\documentclass[sigconf, nonacm]{acmart}

\usepackage{amsmath} 
\usepackage{balance} 
\usepackage{booktabs} 
\usepackage{caption}
\usepackage{comment}
\usepackage{fancyvrb} 
\usepackage[symbol]{footmisc}
\usepackage{graphicx} 
\usepackage{makecell} 
\usepackage{multirow} 
\usepackage{subfigure} 
\usepackage{soul}
\usepackage{xcolor}
\usepackage{xspace} 
\usepackage{enumitem}
\usepackage{amsthm}

\def\BibTeX{{\rm B\kern-.05em{\sc i\kern-.025em b}\kern-.08emT\kern-.1667em\lower.7ex\hbox{E}\kern-.125emX}}

\newcommand{\Execute}{\ensuremath{\mathsf{Execute}}}

\newcommand{\VerifyOp}{\ensuremath{\mathsf{VerifyOp}}}
\newcommand{\VerifyAppend}{\ensuremath{\mathsf{VerifyAppend}}}
\newcommand{\ProveAppend}{\ensuremath{\mathsf{ProveAppend}}}
\newcommand{\digest}{\ensuremath{\mathsf{digest}}}

\newcommand{\remove}[1]{}

\newcommand{\systemname}{\textsc{GlassDB}\xspace}
\newcommand{\op}{\textit{op}}
\newcommand{\verify}{\textit{Verify}}
\newtheorem{definition}{Definition}

\begin{document}

\fancyhead{}
\title{\systemname: Practical Verifiable Ledger Database System Through Transparency}

\settopmatter{authorsperrow=4}
\author{Cong Yue}
\affiliation{
    \institution{National University of Singapore}
    \country{}}
\email{yuecong@comp.nus.edu.sg}

\author{Tien Tuan Anh Dinh}
\affiliation{
    \institution{Deakin University}
    \country{}}
\email{anh.dinh@deakin.edu.au}

\author{Zhongle Xie}
\affiliation{
    \institution{Zhejiang University}
    \country{}}
\email{xiezl@zju.edu.cn}

\author{Meihui Zhang}
\affiliation{
    \institution{Beijing Institute of Technology}
    \country{}}
\email{meihuizhang@bit.edu.cn}

\author{Gang Chen}
\affiliation{
    \institution{Zhejiang University}
    \country{}}
\email{cg@zju.edu.cn}

\author{Beng Chin Ooi}
\affiliation{
    \institution{National University of Singapore}
    \country{}}
\email{ooibc@comp.nus.edu.sg}

\author{Xiaokui Xiao}
\affiliation{
    \institution{National University of Singapore}
    \country{}}
\email{xkxiao@nus.edu.sg}

\sloppy

\begin{abstract}
Verifiable ledger databases protect data history against malicious tampering. Existing systems, such as
blockchains and certificate transparency, are based on transparency logs --- a simple abstraction allowing
users to verify that a log maintained by an untrusted server is append-only. They expose a simple key-value
interface without transactions. Building a practical database from transparency logs, on the other hand, remains a challenge.

In this paper, we explore the design space of verifiable ledger databases along three
dimensions: abstraction, threat model, and performance. We survey existing systems and identify their
two limitations, namely, the lack of transaction support and the inferior efficiency. We then present \systemname, a distributed database system that
addresses these limitations under a practical threat model. \systemname inherits the verifiability of transparency logs, but supports transactions and offers high performance. It extends a ledger-like key-value store with a data structure for
efficient proofs, and adds a concurrency control mechanism for transactions. \systemname batches
independent operations from concurrent transactions when updating the core data structures. In addition, we design
a new benchmark for evaluating verifiable ledger databases, by extending YCSB and TPC-C benchmarks. Using this benchmark, we compare
\systemname against four baselines:
reimplemented versions of three
verifiable databases, and a verifiable map backed by a transparency log. Experimental results demonstrate that \systemname is an efficient, transactional, and verifiable ledger database system. 
\end{abstract}

\maketitle

\section{Introduction}
A verifiable database protects the integrity of user data and query execution on untrusted database providers.
Until recently, the focus has been on protecting the integrity of query execution~\cite{integridb, ads, arx}. In
this context, users upload the data to an untrusted provider which executes queries and returns proofs that certify
the correctness of the results. However, such OLAP-style verifiable databases rely on complex cryptographic primitives
that limit the performance or the range of possible queries.

We observe a renewed interest in verifiable databases, with a focus on OLTP-style systems. In particular, there
emerges a new class of systems, called {\em verifiable ledger databases}, whose goal is to protect the integrity
of the data history. In particular, the data is maintained by an untrusted provider that executes read
and update queries. The provider produces integrity proofs about the data content and its entire evolution history.

An example of verifiable ledger databases is the blockchain~\cite{hyperledger, ethereum, quorum, ruan2023blockchains}. The blockchain
maintains a replicated append-only log in a decentralized setting. It protects the integrity of the log against
Byzantine attackers, by running a distributed consensus protocol among the participants. The integrity proof
(in a permissioned blockchain) consists of signed statements from a number of participants. Another example is
a certificate transparency log~\cite{ct,coniks}, in which a centralized server maintains a tamper-evident,
append-only log of public key certificates. The server regularly publishes summaries of the log which are
then checked for consistency by a set of trusted {\em auditors}. The integrity proof generated by the server
can be verified against the published and audited summaries. The third example is Amazon's
Quantum Ledger Database (QLDB) service~\cite{qldb}, which maintains an append-only log similar to that of
certificate transparency. QLDB uses the log to record data operations that are then applied to another backend database.

Our goal is to build a practical verifiable ledger database system. We observe that the three examples above are built from a common abstraction, namely a {\em transparency log}, which provides two important security properties.
First, users can verify that the log is append-only, namely, any successful update operations will not be
reverted. Second, users can verify that the log is linear, that is, there is no fork in history.
Blockchains enforce these properties by replicating the log and running consensus protocol among the
participants. Certificate transparency log and QLDB rely on auditors or users to detect violations of
the properties. Despite useful security properties, transparency logs are inadequate as databases. In
fact, we identify three challenges in building a practical verifiable ledger database system on top of this abstraction.

The first challenge is the lack of a unified framework for comparing verifiable ledger databases. In
particular, we note that the three systems above have roots from three distinct fields of computer science: blockchains are from
distributed computing, certificate transparency is from security, and QLDB is from database. As a consequence,
there is no framework within which they can be compared fairly.
The second challenge is the lack of traditional database
abstraction, that is, transactions. The transparency logs used in existing systems expose simple key-value interfaces without transactions. This simplifies the design of the transparency logs, but makes them unsuitable for OLTP workloads.
The third challenge is how to achieve high
performance while retaining security. Blockchains, for instance, suffer from poor performance due to the
consensus bottleneck. Certificate transparency has low performance because of expensive disk-based operations, while
QLDB generates inefficient integrity proofs for verifying the latest data.

We address the first challenge by establishing the design space of verifiable ledger databases. The space
consists of three dimensions. The abstraction dimension captures the interface exposed to the users, which
can be either key-value or general transactions. The threat model dimension includes different security assumptions. The performance dimension includes design choices that affect the integrity proof sizes and the overall
throughput. In addition to the design space, we propose a benchmark for comparing the performance of different
verifiable ledger databases. Specifically, we extend traditional database benchmarks, namely YCSB and TPC-C,
with additional workloads containing verification requests on the latest or historical data.

We address the second and third challenges by designing and implementing 
\systemname, a new distributed verifiable ledger database system that overcomes the limitations of
existing systems. \systemname supports distributed transactions and has
efficient proof sizes. It relies on auditing and user gossiping for security.
It achieves high throughput by building on top of a novel data structure: a
two-level Merkle-like tree. This data structure protects the data indexes, which
enables secure and efficient verification. Furthermore, it is built over the
states, as opposed to over transactions, which enables efficient lookup and
proof generation while reducing the storage overhead. \systemname partitions data
over multiple nodes, where each node maintains a separate ledger, and uses the classic 
two-phase commit protocol to achieve transaction semantics. Each node of \systemname
has multiple threads for processing transactions and generating proofs in parallel,
and a single thread for updating the ledger storage. \systemname uses 
optimistic concurrency control to resolve conflicts in transactions, and batching 
to reduce the cost of updating and persisting the core data structure. We conduct an
extensive evaluation of \systemname, and benchmark
it against four baselines: reimplemented versions of three
verifiable databases, and a key-value store based on the transparency log.

In summary, we make the following contributions.
\begin{itemize}[leftmargin=*]
\item We present the design space of verifiable ledger databases, consisting of three dimensions: abstraction,
threat model, and performance. We discuss how existing systems fit into this design space.

\item We design and implement \systemname, a distributed verifiable ledger database system that addresses the limitations of existing works.  In
particular, \systemname supports distributed transactions with high performance, under a practical threat model. 

\item We design new benchmarks for evaluating and comparing verifiable ledger databases. The benchmarks extend YCSB
and TPC-C with workloads that stress test the performance of proof generation and historical data access.

\item We conduct detailed performance analysis of \systemname, and compare it against four baselines, 
namely QLDB~\cite{qldb}, LedgerDB~\cite{ledgerdb}, SQL Ledger~\cite{sqlledger}, and a key-value store based on transparency log, Trillian~\cite{trillian}.  The results show that \systemname consistently outperforms the four baselines across all workloads.
\end{itemize}


\section{Verifiable Ledger Databases}
\label{sec:vldb}



\subsection{What A Verifiable Ledger Database Is}

A verifiable database is a database that ensures the integrity of both the data and query execution. It is useful in outsourced settings, in which the data is managed by an untrusted third party whose misbehavior can be
reliably detected. Until recently,  verifiable databases have focused on ensuring the integrity of query execution,
particularly on the performance and expressiveness of analytical queries that can be verified~\cite{integridb,
ads,arx}.

A verifiable ledger database~\cite{qldb, ledgerdb, sqlledger, spitz} is one instance of the verifiable database, which focuses on protecting the integrity of both the data content and data history. \remove{Figure~\ref{fig:vdb}(a) illustrates how it works.} A user issues a read or
update operation (OLTP query) to the database server (or server), which then executes the operation and appends it to a history log
$H$. The database returns integrity proofs showing that (1) the operation is
executed correctly on the states derived from $H$, and (2) the operation is
appended to $H$, and $H$ is append-only. These proofs ensure that malicious tampering such as changing the data content,
back-dating operations, forking the history log, are detected. Existing works on authenticated data
structure~\cite{ads}, for example, only meet the first condition.

\remove{
\begin{definition}
Let $\op$ be a user query, $H$ be a sequence of operations, i.e., $H=\langle \op_1, \op_2...\rangle$. A
database $D$ executes $\op$ and returns a proof, which is a tuple $\langle \sigma(H), \pi \rangle$ for some function $\sigma$. Let $S(H)$ be
the states derived by applying operations in $H$ in sequence to the initial database states. The user verifies the proof with the function
$\verify(\sigma(H), \op, \pi)$ which returns True if the user accepts the proof, and False otherwise. $D$ is a verifiable ledger database if the following conditions are true:
\begin{itemize}[topsep=2pt,itemsep=1pt,parsep=0pt,partopsep=0pt,leftmargin=11pt]
\item For any $H, \op, \pi$, there does not exist $\op' \neq \op$ such that
$S(H + \op) \neq S(H + \op')$ and $\verify(\sigma(H), \op, \pi) = \verify(\sigma(H), \op', \pi) = \text{True}$

\item For any $H, \op, \pi, H', \op', \pi'$ such that $\verify(\sigma(H), \op, \pi) = \verify(\sigma(H'), \op', \pi') = \text{True}$ and
$|H| \leq |H'|$, then $H$ is a prefix of $H'$.
\end{itemize}
\label{lab:def}
\end{definition}
}

More formally, a verifiable ledger database $D$ consists of four main operations.
\begin{itemize}[leftmargin=*]
        \item $(\digest_{S',H'}, \pi, R, S', H') \leftarrow \Execute(S,H,\op)$: this is
                run by the database server. It takes as input the current state
                $S$ and history log $H$, and the user operation $\op$. It
                executes $\op$, updates the states $S$ accordingly, and appends
                $\op$ to $H$. It returns the updated states $S'$, updated
                history $H'$, a digest value $\digest_{S',H'}$ computed over the new
                state and history, an execution result $R$, and a proof $\pi$.
        \item $\pi \leftarrow \ProveAppend(\digest_{S,H}, \digest_{S',H'})$:
                this is run by the server. It takes as input two digest values
                corresponding to two different history logs. It returns a proof
                $\pi$.
        \item $\{0,1\} \leftarrow \VerifyOp(\op, R, \pi, \digest_{S,H}, \digest_{S',H'})$:
                this is run by the user. It takes as input the user operation
                $\op$, the execution result $R$, the proof $\pi$, and $\digest_{S,H}, \digest_{S',H'}$ that
                correspond to the history and state before and after $\op$ is executed.
                It returns $1$ if the proof is valid, and $0$ otherwise.
        \item $\{0,1\} \leftarrow \VerifyAppend(\digest_{S,H}, \digest_{S',H'},
                \pi)$: this is run by the user. It takes as input two digest
                values corresponding to two different history logs, and a
                proof $\pi$. It returns $1$ if the proof is valid, and $0$
                otherwise.
\end{itemize}

\begin{definition}
A verifiable ledger database, which supports the four operations defined above, is secure if it satisfies the following properties.
\begin{itemize}[leftmargin=*]
    \item {\bf Integrity:} the database server cannot tamper with the user operation without being detected. More precisely, given any $\op, \pi, R, S,S', H, H', \digest_{S,H}, \digest_{S',H'}$ such that $(\digest_{S',H'},\pi,R,\_,\_) \leftarrow \Execute(S,H,\op)$, $\VerifyOp(\op,R,\pi, \digest_{S,H},\digest_{S',H'}) = 1$, the server cannot find $\pi', R'$ such that $R' \neq R$ and $\VerifyOp(\op,R',\pi',\digest_{S,H},\digest_{S',H'}) = 1$.
    \item {\bf Append-only:} the database server cannot fork the history log without being detected. More precisely, for any $\op, S_1, H_1, S, H, \digest_{S,H}, \digest_{S_1,H_1}, R, \pi$ such that $(\digest_{S_1,H_1},\pi,R,S_1,H_1) \leftarrow \Execute(S,H,\op)$ and $\VerifyOp(\op,R,\pi,\digest_{S,H}, \digest_{S_1,H_1})=1$, the server cannot find $\op', S_2, H_2, S', H', \digest_{S',H'}, \digest_{S_2,H_2}, R', \pi', \pi_a$  such that $(\digest_{S_2,H_2},\pi',R',S_2,H_2) \leftarrow \Execute(S',\ H',\ \op')$, $\VerifyOp(\op',\ R',\ \pi',\ \digest_{S',H'},\ \digest_{S_2,H_2})=1$, $\pi_a \leftarrow \ProveAppend(\digest_{S_1,H_1}, \digest_{S_2,H_2})$, $\VerifyAppend(\digest_{S_1,H_1}, \digest_{S_2,H_2}, \pi_a) = 1$, $|H_1| < |H_2|$, and $H_1$ is not a prefix of $H_2$.
\end{itemize}
\label{lab:def}
\end{definition}

\remove{
\textit{Example.} We use Figure~\ref{fig:vdb}(b) to illustrate the properties of verifiable ledger
databases. Suppose the current history at database $D$ is $H_1$, corresponding to the states $\{a=1, b=2\}$.
When a user submits a query that adds value $c=3$, the states are updated to $\{a=1, b=2, c=3\}$, and the
operation $\textit{add(c,3)}$ is appended to the $H_1$. $D$ returns  a proof $\langle \sigma(H_1), \pi\rangle$. The
user verifies it with $\verify(\sigma(H_1),\textit{add(c,3)}, \pi)$. The first condition of Definition~\ref{lab:def}
means that $\pi$ captures the states $\{a=1, b=2, c=3\}$, such that if the
verification passes, any operation $\op \neq \textit{add(c,3)}$ will make $\verify(\sigma(H_1), \op, \pi)$ fails. In
other words, if $D$ updates the states to $\{a=1, b=2, c=4\}$, for example by replacing user's operation
with $\textit{add(c,4)}$, the verification will fail.

Figure~\ref{fig:vdb}(c) illustrates another scenario, in which starting from the same $H_1$, one user issues a sequence of
queries that adds $c=3$, removes $c$, then adds $d=4$. The database returns $(\sigma(H_3, \pi))$ for the last operation,
and $\verify(\sigma(H_3), \textit{add(d,4)}, \pi)$ succeeds. The second condition of Definition~\ref{lab:def} forbids
the case when another user adds $d=4$ and the database returns $\sigma(H_1), \pi'$ such that
$\verify(\sigma(H_1), \textit{add(d,4)}, \pi')$ also succeeds. In other words, the database cannot fork the
history, even if the two histories result in the same states. Otherwise, $D$ can remove operations from its
history, for instances the two operations on $c$.
}

\textit{Example.} \remove{We use Figure~\ref{fig:vdb}(b) to illustrate the
properties of verifiable ledger databases.} Suppose the current history at the 
database $D$ is $H_1$, corresponding to the state $S_1 = \{a=1,
b=2\}$.  
Starting from
the same $H_1$, one user $A$ issues a sequence of operations $\langle
\mathsf{add}(c,3), \mathsf{remove}(c), \mathsf{add}(d,4) \rangle$, while
another user $B$ issues $\mathsf{add(d,4)}$. This scenario can happen due to
concurrency, or because the server acts maliciously.  For user $A$'s last
operation, the server returns $\digest_{S_4,H_4}, \pi_4, R_4$. For user $B$'s
operation, it returns $\digest_{S_2',H_2'}, \pi_2', R_2'$.  The append-only property
means that for $\pi \leftarrow \ProveAppend(\digest_{S_2',H_2'},
\digest_{S_4,H_4})$, $\VerifyAppend(\digest_{S_2',H_2'}, \digest_{S_4,H_4},
\pi) = 0$. In other words, the server cannot fork the history, even if the two branches result in the same state.

\begin{table*}
\footnotesize
\centering
\caption{\systemname vs. other verifiable ledger databases. N is number of transactions, m is number of keys, and B is number of blocks, where $m \ge N \ge B$. All systems, except Forkbase, support inclusion proof of the same size as append-only proof.}
\vspace{-2mm}
\resizebox{\textwidth}{!}{\begin{tabular}{p{3.5cm}|c|c|c|c|c|c}
\hline 
{\bf System} & {\bf Data Model} & {\bf Transaction} & {\bf Threat model} & {\bf Append-Only Proof} & {\bf Current-Value Proof} & {\bf Throughput} \\
\hline 
QLDB~\cite{qldb} & Relational & Transaction & Audit & $O(\log N)$ & $O(N)$ & Low \\ \hline
LedgerDB~\cite{ledgerdb} & Key-value & Transaction & Audit & $O(\log N)$ & $O(N)$ & Medium \\ \hline
SQL Ledger~\cite{sqlledger} & Relational & Transaction & Audit & $O(B)$ & $O(N)$ & Medium \\ \hline
Forkbase~\cite{forkbase} & Key-value & Non-transaction & Audit & $O(N)$ & $O(\log m)$ & Medium \\ \hline
Blockchain~\cite{hyperledger} & Key-value & Transaction & Consensus & $O(1)$ & $O(1)$ & Low \\ \hline
CreDB~\cite{credb} & Key-value & Transaction & Trusted hardware & $O(1)$ & $O(1)$ & Low \\ \hline
Trillian~\cite{trillian}, ECT~\cite{ect},  $\text{Merkle}^2$\cite{merkle2}& Key-value & Non-transaction & Audit & $O(\log m)$ & $O(\log m)$ & Low  \\ \hline
{\bf \systemname} & Key-value & Transaction & Audit & $O(\log B)$ & $O(\log B + \log m)$ & High \\ \hline
\end{tabular}}
\vspace{-2mm}
\label{tab:overview}
\end{table*}

\subsection{Design Space}
\label{sec:security-definition}
Definition~\ref{lab:def} admits a simple, naive design in which the proof consists of the query result and
complete history $H$ (signed with the provider's cryptographic key). The users replay all operations in $H$ to
verify the correctness of $H$, and they broadcast messages among each other to detect any inconsistent behavior, e.g., the database signed different histories that were not linear. However, this design incurs
significant communication and computation costs for the users. A more practical design would need to reduce these
costs. To enable a principled comparison of different verifiable ledger databases, we propose to explore the design space along three dimensions: abstraction, threat model, and performance. 

\subsubsection{Abstraction.} This refers to the data model and programming model supported by the database.
There are two main data models with different trade-offs. On the one hand, the key-value model exposing simple {\tt Put} and {\tt Get} operations is flexible and scalable. On the other hand, the relational data model supports declarative query languages and is easier to use. The key-value data model is more suitable for ledger databases running OLTP workloads since it simplifies the verification logic. In contrast, the relational model entails more complexity, because it uses secondary indexes and supports complex operations such as join and aggregation, which are difficult to verify. Systems such as SQL Ledger and QLDB support the relational model, but they do not guarantee the integrity of indexes and operations.

The two main programming models in ledger databases are transactional and non-transactional. In the former, users can execute multiple operations in one transaction, with serializable properties (ACID). Examples include QLDB, LedgerDB, and SQL Ledger. In the latter, the database performs one operation at a time, without guarantees across multiple operations. Examples include systems such as Trillian, Merkle$^2$, and Coniks.
There are other design choices between non-transactional abstraction and ACID transactional abstraction. In particular, some database
systems support serializable transactions over small sets of related
keys~\cite{bigtable,gstore}, or keys within the same partitions~\cite{hstore}. Some other
databases support transactions with weaker isolation levels, such as snapshot
isolation~\cite{percolator}. These design choices can deliver higher
performance than the design with serializable transactions, but they suffer from
anomalies. We note that in the context of verifiable ledger databases, such anomalies can
happen due to the server acting maliciously to cause conflict during execution,
instead of due to real concurrency. As a consequence, the application needs to
handle a potentially large number of anomalies, which increases complexity and
performance overhead. For the rest of the paper, we use the term transactions
to refer to serializable transactions.

\subsubsection{Threat model.} The security of a verifiable ledger database is defined as having integrity proofs
that satisfy the two conditions in Definition~\ref{lab:def} under a specific threat model. All threat
models share common assumptions that the attacker cannot break cryptographic primitives or mount denial of
service attacks.   

There are three main threat models in the context of verifiable ledger databases. The most common model involves
a single untrusted database provider that behaves in a Byzantine manner. It has been shown that in this
setting, it is only possible to achieve {\em fork consistency}~\cite{sundr}, i.e., users cannot prevent misbehavior but can only detect it by communicating with each other. As a result, this model assumes
that users engage in gossiping, and that the attacker cannot permanently partition the network. To further
reduce the cost on the users, the model can be extended by introducing a set of trusted, powerful users called
{\em auditors} that only gossip among themselves. The auditors check for the misbehavior of the database on behalf
of the users. 

The second threat model assumes that the database is replicated over a set of providers, the majority of which
are trusted. Even though there are malicious providers, the system as a whole enforces the correct behavior.
In particular, the providers participate in a distributed, Byzantine fault tolerant consensus protocol to
ensure consistency of the database~\cite{pbft}. We note that such consensus-based systems provide stronger
security guarantees than systems with single malicious providers, that is they can prevent misbehavior as
opposed to only detecting it.  

The final threat model assumes that the database server is malicious, but it is
equipped with some trusted hardware that supports trusted execution environments
(TEEs). The TEE protects the computation and data running inside the environment against
malicious operating systems and hardware attacks. The entire database can run
securely inside the TEE. However, this model assumes that both the computation
and the TEE itself are free of vulnerabilities, which
does not always hold in practice~\cite{foreshadow}.

\subsubsection{Performance.} The performance of a verifiable ledger database is evaluated based on two
metrics: the user's verification cost, and the database throughput. The former depends on the complexity of
the integrity proofs. An efficient proof is short and fast to verify. We further categorize integrity proofs
into three types. 

\begin{itemize}[leftmargin=*]
\item Inclusion proof: given $\digest_{S,H}$ and a value $v$ corresponding to a key $k$, this proof ensures that $v$ is included at some point in $H$.

\item Current-value proof: given $\digest_{S,H}$ and a value $v$ of a key $k$, this proof ensures that $v$ is the latest value of $k$ in $H$.

\item Append-only proof: given $\digest_{S,H}$ and $\digest_{S',H'}$, the proof ensures that $H$ is a prefix of $H'$
(assuming that $|H| \leq |H'|$).
\end{itemize}

The database throughput is measured in terms of the number of user queries completed per second, and it depends on
the cost of maintaining the security-related data structures for generating the proofs. Designs that exploit
parallel execution and avoid contention will have high throughputs.

\remove{
\subsection{Transparency Log}
\label{subsec:proof}
Transparency log is an append-only log accompanied by a Merkle
tree. Figure~\ref{fig:merkle} shows an example of a transparency log at two different times: one with $3$
elements and another with $6$ elements. Each leaf represents an operation, for example updating a key-value
tuple. The proof for an append operation is the new Merkle root tree.  
\begin{itemize}[leftmargin=*]
\item Inclusion proof:  the proof is the Merkle path from the
corresponding leaf to the root. The cost of this proof is $O(\log(N))$ where $N$ is the size of the log. In the
example, the proof that element 2 exists in the log consists of the hashes of node $1$ and $b$. 
\item Append-only proof: the proof consists of nodes for reconstructing both trees, which has the complexity of $O(\log(N))$. In the example, to
prove that $c$ is included in $l$, the proof includes the hashes of node $e, 3, 4, k$. The first three are
sufficient to compute $c$, and all four are sufficient to compute $l$.  
\item Current-value proof: the proof
includes all the leaves of the tree,  which has the complexity of $O(N)$. In our
example, suppose the latest value for key $k$ is set at node $3$. Given $l$ and the tuple, the user has to fetch all $6$
elements and check that nodes $4, 5, 6$ do not update the tuple.  
\end{itemize}
The security of transparency logs depends on auditing. In particulars, users broadcast signed Merkle roots to a
number of auditors which check that that there is a single log with no forks. The check is done by requesting
and verifying append-only proof from the database provider. 

Transparency log is an append-only log protected by a Merkle
tree~\cite{crosby09}.
Each leaf represents an operation, for example updating a key-value
tuple. This data structure supports all three types of integrity proofs. The {\em inclusion proof} consists  of the Merkle path from the
leaf to the root, which costs $O(\log(N))$ where $N$ is the size of the log. 
The {\em append-only proof} include intermediate nodes between two trees, and has the cost of $O(\log(N))$. The {\em current-value proof}, however, requires all the leaves of the tree, thus its cost is $O(N)$.

The security of transparency logs depends on auditing. In particulars, users broadcast signed Merkle roots to a
number of auditors which check that that there is a single log with no forks. The check is done by requesting
and verifying append-only proof from the database provider.
}

\subsection{Review of Existing Systems}
\label{subsec:review}
Table~\ref{tab:overview} compares existing verifiable ledger databases according to the design space above. These
systems build on top of the transparency logs described above. 

\textit{Transparency logs.} 
Transparency log is an append-only log protected by a Merkle
tree~\cite{crosby09}. Each leaf represents an operation. 
This data structure supports all three types of integrity proofs. The {\em inclusion proof} consists of the Merkle path from the
leaf to the root, which costs $O(\log(N))$ where $N$ is the size of the log. 
The {\em append-only proof} includes intermediate nodes between two trees, and has the cost of $O(\log(N))$. The {\em current-value proof}, however, requires all the leaves of the tree, thus its cost is $O(N)$. Trillian~\cite{trillian} combines transparency logs and maps to implement new
primitives called {\em verifiable log-based map} for storing public key certificates. When a key is updated, the
map is updated and a new Merkle root is computed on the map. It then appends the log with both the operation
and the Merkle root. As the result, both the current-value and append-only proofs are efficient, i.e.
$O(\log(m))$ complexity. Other systems CONIKS\cite{coniks}, ECT~\cite{ect}, and $\text{Merkle}^2$~\cite{merkle2} improve Trillian by adding support for privacy, revocation (non-inclusion proofs), and reducing the audit cost. 

\textit{Commercial verifiable ledger databases.} QLDB~\cite{qldb}, LedgerDB \cite{ledgerdb}, and SQL Ledger \cite{sqlledger} are recent services offered by major cloud providers. QLDB uses transparency logs for storing transactions,
and executes the operations on indexed tables. However, its throughput is low due to the disk-based
communication between the log and the indexed tables. LedgerDB~\cite{ledgerdb} and SQL Ledger \cite{sqlledger} improve the performance of QLDB by batching multiple
transactions when updating the Merkle roots of the log. 
Since QLDB and LedgerDB build Merkle trees over transactions, the append-only proof costs $O(\log(N))$. However, SQL Ledger store the blocks in a hash chain, requiring $O(B)$ cost for append-only proof. All three systems do not have protection over indexes, which require scanning to the latest transaction for current-value proof. The cost of this is $O(N)$.

\textit{Forkbase.} Forkbase~\cite{forkbase} is a state-of-the-art versioned, key-value storage system.
It implements a variant of transparency logs called transparency maps. In particular, Forkbase builds a Merkle
tree on top of immutable maps: each update operation results in a new map and a new Merkle root. Each Merkle
root also includes a cryptographic pointer to the previous root. Unlike transparency logs, the current-value
proof in Forkbase costs $O(\log(m))$ because the latest value is included in the map. However, the append-only
proof is $O(N)$, since users have to follow the hash chain to ensure that there are no forks.  

\textit{Blockchain.} Existing blockchain systems assume the majority of trusted providers in a
decentralized setting. The providers run a Byzantine fault tolerant consensus protocol to keep the ledger and
global states consistent. As the system as a whole is trusted, signed statements from the blockchain can be
used as integrity proofs. In particular, in a permissioned blockchain that tolerates $f$ Byzantine failures,
the proof contains signatures from $f+1$ providers. The proof complexity is therefore independent of the
history, or $O(1)$. However, the performance of a blockchain is limited by the consensus
protocol~\cite{blockbench}.  

\textit{CreDB} Instead of relying on consensus to protect against
Byzantine database servers, CreDB assumes that the server can create trusted
execution environments backed by trusted hardware. CreDB processes user
transactions inside the TEE and produces signed {\em witnesses} that capture the
history of the states. Both inclusion and current-value proofs in CreDB are efficient, because they
are simple messages signed by the TEE. However, the hardware limitation, e.g. the limited memory available for Intel SGX, makes the TEE a
performance bottleneck, and results in low throughputs.

\remove{
\textit{Public-key transparency logs.} Trillian~\cite{trillian} combines transparency logs and maps to implement new
primitives called {\em verifiable log-based map}. It exposes a key-value interface, and is used for storing public key certificates. When a key is updated, the
map is updated and a new Merkle root is computed on the map. It then appends the log with both the operation
and the Merkle root. As the result, both the current-value and append-only proofs are efficient, i.e.
$O(\log(m))$ complexity. On the other hand, Trillian relies on trusted auditors to regularly verify the
consistency between the log and the corresponding map. More specifically, the auditors reconstruct the map
from the log and ensure that the Merkle root included in the log is correct. This auditing process is
expensive and should be performed by powerful, external entities rather than by the users. Other systems such as CONIKS\cite{coniks}, ECT~\cite{ect}, and $\text{Merkle}^2$~\cite{merkle2} are similar to Trillian in our design space. They improve Trillian by adding support for privacy, revocation (non-inclusion proofs), and reducing the audit cost. 
}


\section{\systemname}
\label{sec:design}

\subsection{Existing Designs}
\label{subsec:existingdesign}

\begin{figure*}
\begin{minipage}{0.48\textwidth}
    \centering
    \includegraphics[height=3.2cm]{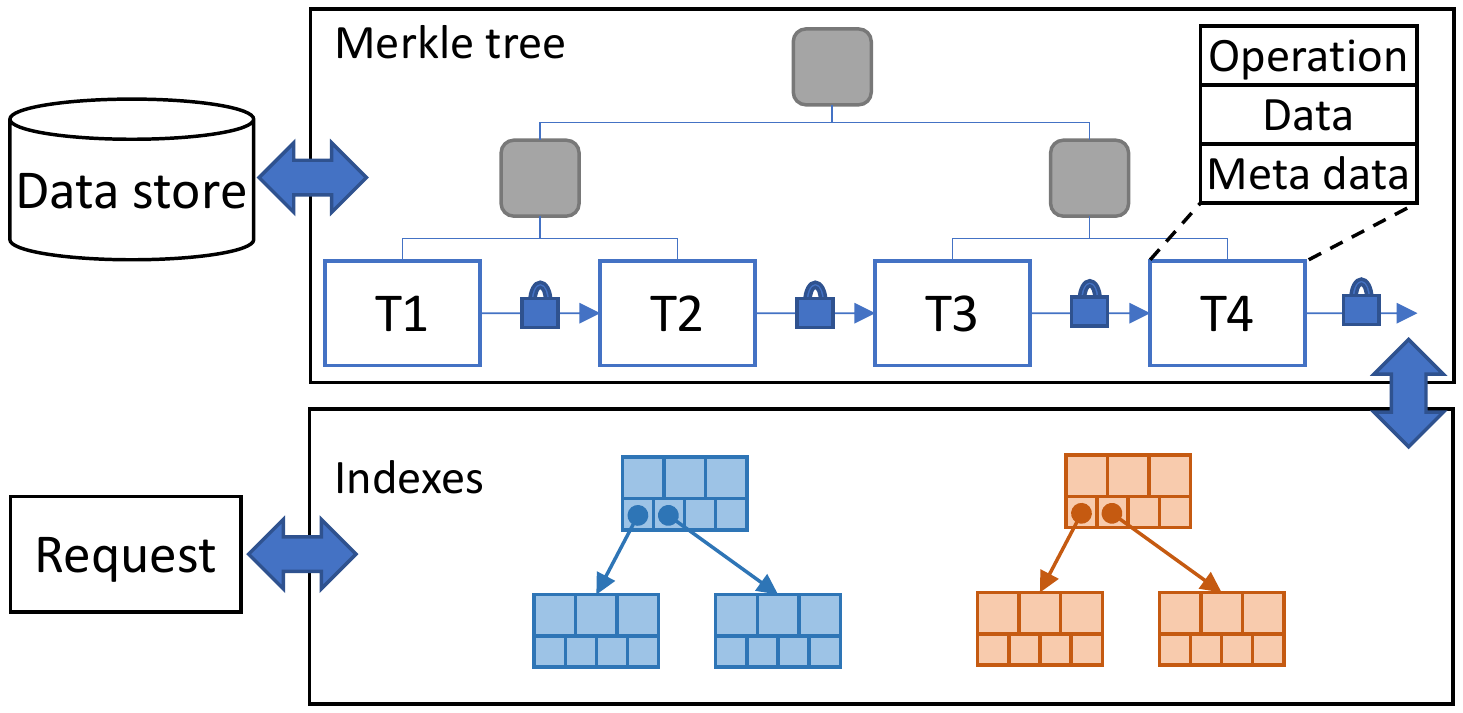}
    \caption{A simple verifiable ledger database. The ledger consists of blocks containing the operation, data, and metadata of the transactions.}
    \label{fig:qldb}
\end{minipage}
\hspace{0.03\textwidth}
\begin{minipage}{0.48\textwidth}
    \centering
    \includegraphics[height=3.2cm]{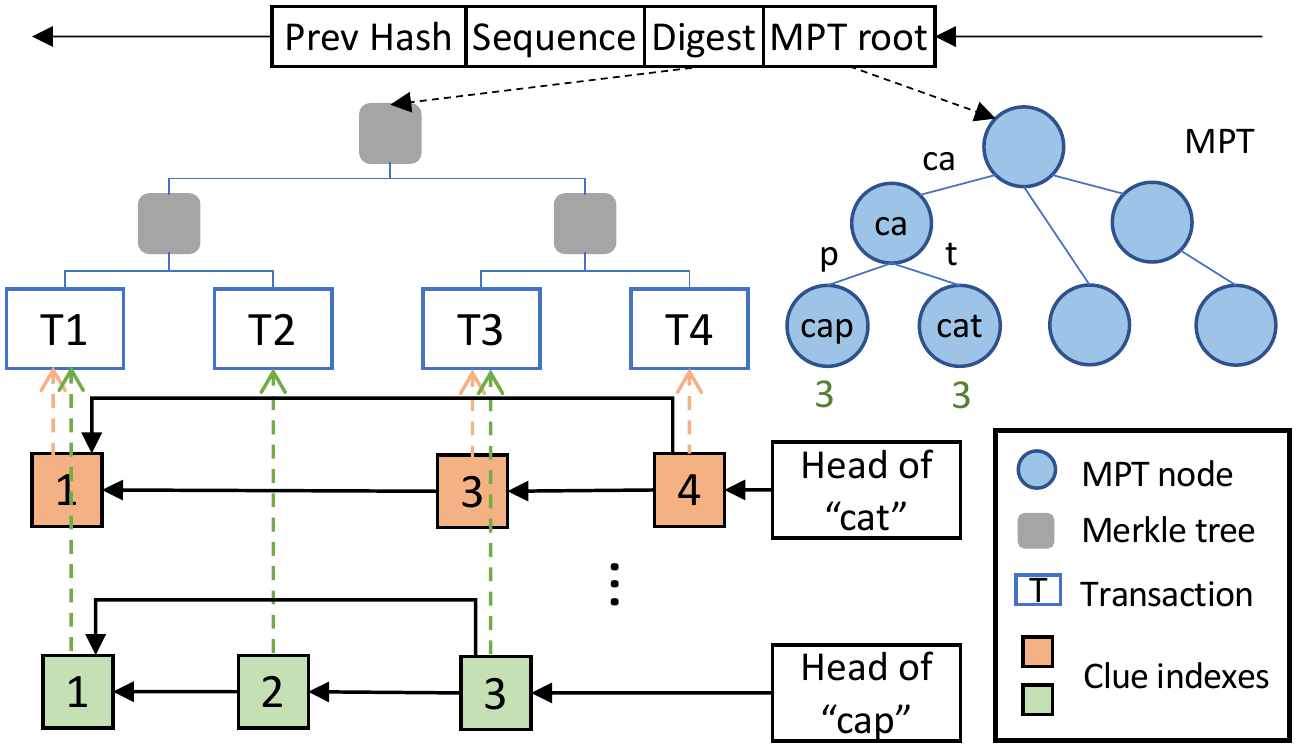}
    \caption{LedgerDB~\cite{ledgerdb} authenticated data structures. There is one clue index for every key, and the size of each clue index is stored in a leaf of the Merkle Patricia Trie.}
    \label{fig:ledgerdb}
\end{minipage}
\vspace{-2mm}
\end{figure*}

Figure~\ref{fig:qldb} shows a design of verifiable ledger databases used in commercial
systems such as QLDB~\cite{qldb}.
The key idea is to replace the transaction log in conventional databases with a variant of transparency log
called {\em ledger}.  The ledger is a hash-chained sequence of blocks, each of which contains the operation
type and parameters, and 
a Merkle tree is built on top of them to protect their integrity. 
Updating the ledger requires appending a new transaction block and rebuilding the Merkle tree. 

Transaction execution in this design is similar to that in a conventional database.
The transaction is first committed to the ledger as a new block, under some concurrency
control mechanisms. Then, the data and indexes are updated. 
The transaction is considered committed once the ledger is updated and the data can be queried via the 
indexes.
The response to the client includes a block sequence number indicating where in the ledger the transaction is committed.
During verification, the client requests a digest of the ledger, and then sends a {\tt GetProof}
request containing the sequence number and the digest.
It receives a Merkle proof showing that the specified block is included in the ledger.
After verifying the proof, it checks that the data is included in the block. 



The main advantage of this design is that it is easy to extend an existing database system into a verifiable ledger database. In addition, the design is independent of the underlying data abstraction and layout.
However, it has two limitations. 
First, it incurs significant overhead in transaction processing, because updates of the Merkle tree are in the critical path. Second, the indexes are not integrity protected, i.e., the server can respond with stale data. As a result, this design requires the client to scan the ledger to guarantee the returned value is current, which incurs $O(N)$ cost. 

LedgerDB~\cite{ledgerdb} improves the design above by updating the authenticated data structures asynchronously.
This technique is also adopted by SQL Ledger.
Figure~\ref{fig:ledgerdb} shows the different indexes and Merkle trees used in LedgerDB.
In LedgerDB, each
transaction is appended to a ledger in the form of a journal entry, and a Merkle tree built on top of the ledger is updated asynchronously in batch, which is called batch accumulated Merkle-tree (bAMT). LedgerDB maintains a skip-list index (called a {\em clue index}) for each individual data key, with each entry in the skip list pointing to the journal entry corresponding to the transaction that modifies the data key. The size of the index is stored as a leaf of a Merkle Patricia Trie, called clue-counter MPT(ccMPT). The roots of ccMPT and bAMT are stored as a block in a hashed chain of blocks. LedgerDB also supports data freshness by using another ledger that stores time entries from a timestamp authority.

There are three limitations of LedgerDB's design that result in high verification costs. First, bAMT stores one transaction per leaf, therefore its size can be large when there are many transactions, which leads to larger proofs. Second, it is expensive to verify a value of a key, even in the presence of a trusted auditor. In particular, the ccMPT structure used to protect the clue index is not secure, because each leaf of the ccMPT stores only the size of each clue index, instead of capturing the content of the entire index. As a consequence, to verify the value of a key, the client needs to scan and verify the entire index to ensure that each entry in the clue index points to a correct journal entry. We note that even if a trusted auditor verifies the ccMPT and clue indexes, the client still needs to verify the clue index by itself, because a malicious server can modify the index without changing the ccMPT. Finally, the size of the proof for multiple keys, even when the keys belong to the same transaction, grows linearly with the number of keys because each key requires a separate proof from the ccMPT.


\vspace{-3mm}
\subsection{\systemname Overview}
\label{subsec:design:our}


\systemname is a new, distributed verifiable ledger database system that overcomes the limitations of the existing designs.
\remove{
\st{It achieves high transaction throughputs by updating the authenticated data structure asynchronously and with aggressive batching. It reduces the overhead of verification by using a two-level pattern-oriented split (POS) tree
that protects the integrity of the entire database states. This tree enables the efficient generation of proofs with small sizes. In addition, by storing the entire states at the lower POS-tree, \systemname guarantees data freshness without relying on a trusted time authority, as the client can verify if the data values correspond to the latest block.}
}
It supports general transactions, which makes it easy to use for existing and
future applications. It adopts the same threat model as QLDB and LedgerDB,
which assumes that the database server is untrusted, and there exists a set
of trusted auditors that gossip among each other.  \systemname achieves high
throughputs and small verification costs. Table~\ref{tab:overview} shows how
the system fits in the design space.

There are three novelties in the design of \systemname that facilitate its
high performance.  First, \systemname adopts hash-protected index structures.
The key insight we identify from the limitation of existing ledger databases is
the lack of comprehensive and efficient protection of the indexes, which leads
to either security issues or high verification overhead. Such limitations can
be eliminated by adopting hash-protected index structures.  Second, \systemname
builds its ledger over the state of data instead of transactions.  One advantage of
this approach is that the system can retrieve the data and generate
current-value proofs more efficiently.  Another advantage is that it results in
a smaller data structure.  The Merkle trees of the existing systems are built
over the transactions, which grow quickly and lead to higher storage and
computation overhead. In contrast, \systemname's core data structure grows more
slowly as it batches updates from multiple transactions.  
Third, \systemname
partitions the data over multiple nodes, which enables it to scale to achieve
high throughput. Furthermore, it adopts three optimizations that help speed up
transaction processing and verification, namely transaction batching,
asynchronous persistence, 
and deferred verification.

\begin{figure*}
    \begin{minipage}{0.48\textwidth}
        \centering
        \includegraphics[height=4.2cm]{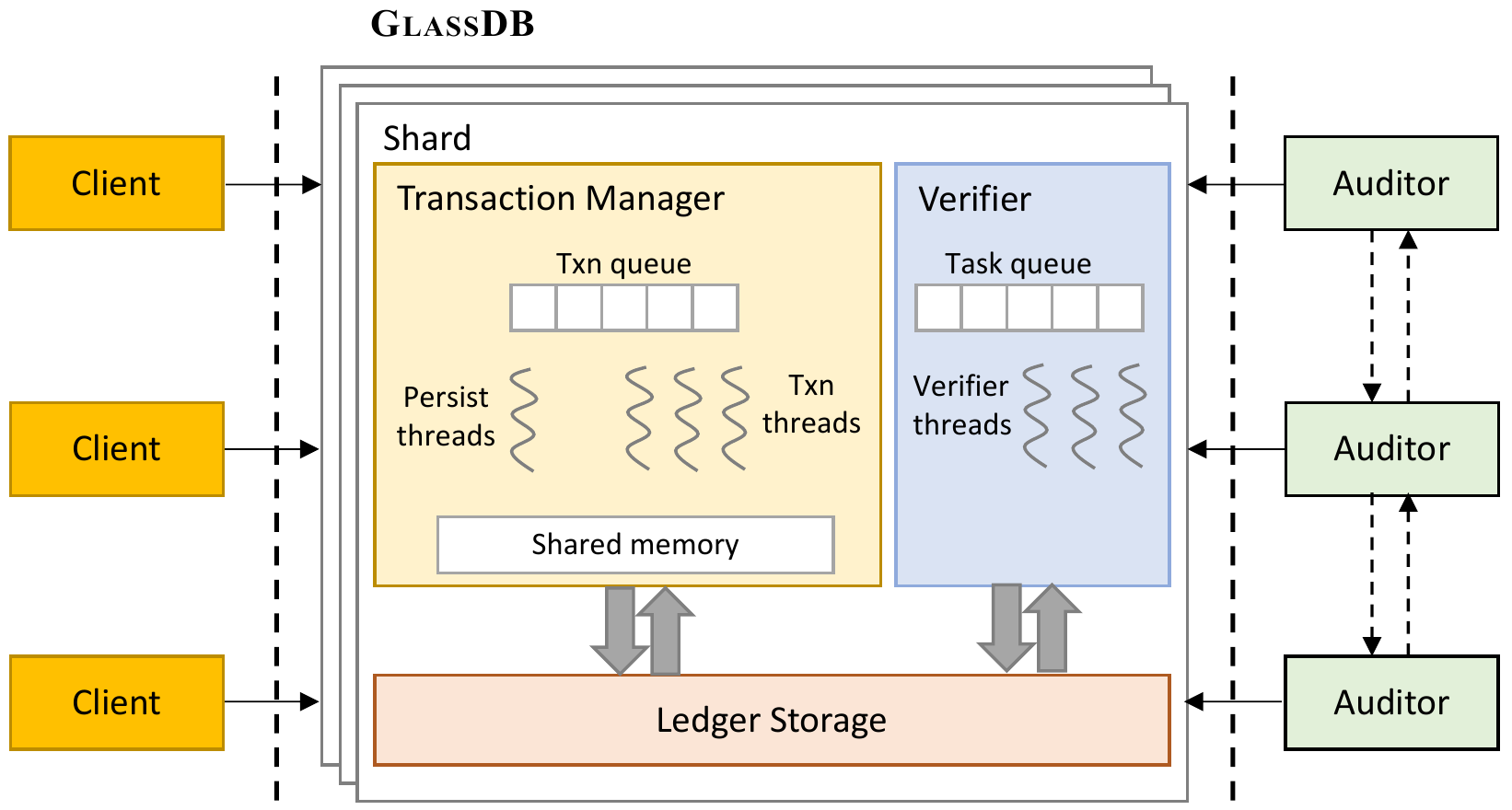}
        \caption{\systemname design. The transaction manager executes transactions, with a persisting thread asynchronously append committed data to the ledger storage, and the verifier handles proof requests.}
        \label{fig:architecture}
    \end{minipage}
    \hspace{0.03\textwidth}
    \begin{minipage}{0.48\textwidth}
        \centering
        \includegraphics[height=4.8cm]{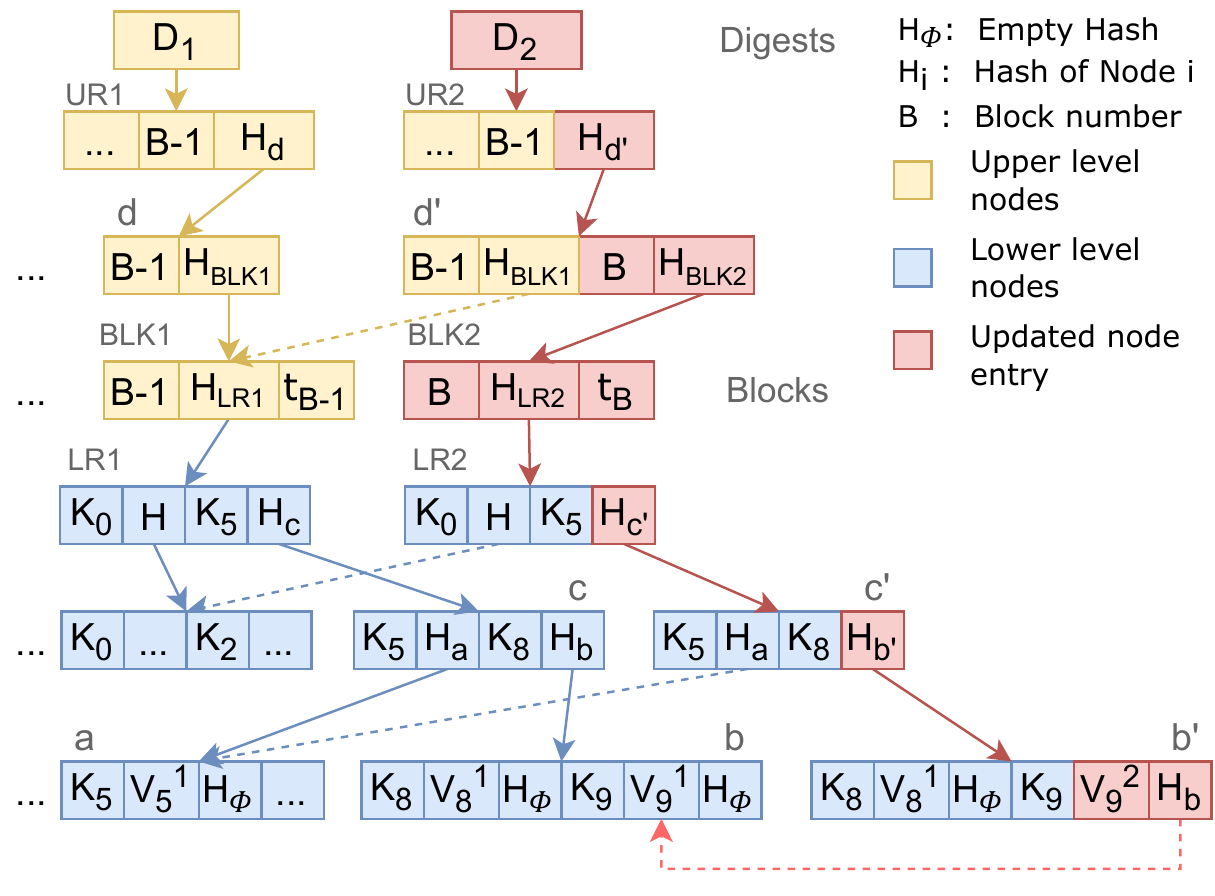}
        \caption{Two-level POS-tree.}
        \label{fig:2lvlpostree}
    \end{minipage}
\vspace{-2mm}
\end{figure*}

Figure~\ref{fig:architecture} shows the design of \systemname. It partitions the data (modeled as key-value tuples) into different shards based on the hash of the keys, and uses two-phase commit (2PC) protocol to ensure the atomicity of cross-shard transactions. Each shard has three main components: a transaction manager, a verifier, and a ledger storage. A transaction request is forwarded to the transaction manager, which executes the transaction using a thread pool with optimistic concurrency control. A verification request is forwarded to the verifier, which returns the proof. 
The ledger storage maintains the core data structure that provides efficient data access and proof generation.
Each shard maintains an individual ledger based on the records committed.
The client keeps track of the key-to-shard mapping, and caches the digests of the shards' ledgers. \systemname uses write-ahead-log (WAL) to handle application failures. It handles node failures by replicating the nodes. 

The life cycle of a transaction at the server can be divided into four phases: prepare, commit, persist, and get-proof.
The prepare phase checks for conflicts between concurrent transactions before making commit or abort decisions.
The commit phase stores the write set in memory and appends the transaction to a WAL for durability and recovery.
The persist phase appends the committed in-memory data to the ledger storage and updates the authenticated data structures for future verification.
The get-proof phase generates the requested proofs for the client. 
In \systemname, the persist and get-proof phases are executed asynchronously and in parallel with the other two phases. The detail is illustrated in Section~\ref{subsec:design:detail}.

\vspace{-3mm}
\subsubsection{APIs}
\label{subsec:api}

\remove{
\noindent
The user (or client) uses the following APIs to interact with the server and the auditor.
\begin{itemize}[leftmargin=*]
    \item {\tt Init(sk)}: initializes the client with private key $sk$ for
    signing the transactions, and sends public key $pk$ to the auditors for
    verification.
    \item {\tt BeginTxn()}: starts a transaction. It returns a unique
            transaction ID $tid$ based on client ID and timestamp.
    \item {\tt Get(tid, key, (timestamp | block\_no))}: returns the latest value of the given key (default option), or the value of the key before the given timestamp or block number.
    \item {\tt Put(tid, key, value)}: buffers the write for the ongoing transaction.
    \item {\tt Commit(tid)}: signs and sends the transaction to the server for commit. It returns a promise.
    \item {\tt Verify(promise)}: requests the server for a proof corresponding to the given promise, then verifies the proof.
    \item {\tt Audit(digest, block\_no)}: sends a digest of a given block to the auditors.
\end{itemize}

The auditors use the following APIs to ensure that the database server is working correctly.
\begin{itemize}[leftmargin=*]
    \item {\tt VerifyBlock(digest, block\_no)}: requests the server for the block at
            $block\_no$, proof of the block, and the signed block transactions.
                It verifies that all the keys in the transactions are included
                in the ledger.
    \item {\tt VerifyDigest(digest, block\_no)}: verifies that the given digest
            and the current digest correspond to a linear history, by asking
                the server to generate append-only proofs. If the given block
                number is larger than the current block number, it uses {\em
                VerifyBlock} to verify all the blocks in between.
    \item {\tt Gossip(digest, block\_no)}: broadcasts the current digest and
            block number to other auditors.
\end{itemize}
}

\noindent
\systemname supports the key-value data model with ACID transactions.
The user (or client) starts by calling {\tt Init(pk, sk)}, which initializes the client's session with the private key $sk$
for signing transactions, and sends the corresponding public key $pk$ to the auditors for
verification. The client invokes {\tt BeginTxn()} to start a transaction, which returns a transaction ID
$tid$ based on the client ID and timestamp.  During the transaction, the client uses {\tt Get(tid, key, (timestamp |
block\_no))} and {\tt Put(tid, key, value)}. When ready to commit, it invokes {\tt Commit(tid)},
which signs and sends the transaction, including the buffered writes,  to the server. This
API returns a promise, which can be passed to {\tt Verify(promise)} to request proof and verify it.
The client frequently invokes {\tt Audit(digest, block\_no)} to send a digest of a given block to the auditors.

The auditor uses {\tt VerifyBlock(digest, block\_no)} to request the server for the block at {\tt block\_no}, proof of the block, and the signed block transactions.
It checks that all the keys in the transactions are included in the ledger.
It uses  {\tt VerifyDigest(digest, block\_no)} to verify that the given digest
and the current digest correspond to a linear history, by asking
the server to generate append-only proofs.
If the given block number is larger than the current block number, it uses {\tt VerifyBlock} to verify all the blocks in between. 
Finally, the auditor calls {\tt Gossip(digest, block\_no)} to broadcast the current digest and block number to other auditors.






\subsection{\systemname Design}
\label{subsec:design:detail}

\subsubsection{Ledger storage}
\label{subsubsec:ledgerstorage}

The design goal of \systemname is to build a storage system that not only offers efficient access to the data, but also
supports efficient inclusion, latest, and append-only proofs. To this end, we use a Merkle variant
called two-level pattern-oriented split tree (or two-level POS-tree).
\remove{
\textit{SIRI} is a class of immutable indexes that combine Merkle trees with the index trees. Each node of the index is identified by the hash of its content. The index node stores the hashes of the child nodes instead of physical addresses. New instances are created from the previous instances using copy-on-write. \textit{SIRI} supports efficient data access by traversing the index, and efficient proofs that contain only neighboring nodes in the path from the leaf to the root. In addition, it is space-efficient because nodes with the same content are reused. In \systemname, we use POS-tree, a type of \textit{SIRI} structure with a high deduplication ratio~\cite{forkbase,siri}.}

A POS-tree is an instance of {Structurally Invariant and Reusable Index} (SIRI)~\cite{forkbase,siri}, which combines the Merkle tree and balanced search tree. A parent node in the POS-tree stores the cryptographic hash of its child nodes, such that the root node contains the digest of the entire tree. The user can perform efficient data lookup by traversing the tree.  The POS-tree is built from the globally sorted sequence of data. The data is split into leaf nodes using content-defined chunking, in which a new node is created when a pattern is matched. The cryptographic hash values of the nodes in one level form the byte sequence for the layer above. The byte sequence is split into index nodes using similar content-defined chunking approach.  POS-tree is optimized for high deduplication rates because of its content-defined chunking. It is immutable, that is, a new tree is created, using copy-on-write,
when a node is updated. Finally, the POS-tree is {\em structurally invariant}, that is, the structure of the tree does not depend on the order of data inserted.



The core data structure of \systemname is shown in  Figure~\ref{fig:2lvlpostree}. It consists of an upper level POS-tree and a lower level POS-tree.
The lower level POS-tree is built on the database states and serves as the index. The leaf nodes store key-value tuples. For each key, the leaf node stores the pointer to the node containing the previous version of the key. For example, the Node $b'$ stores the key $K_9$ with value $V_9^2$, and $H_b$ which is the hash of node $b$ where $V_9^1$ is. 
The internal nodes of this tree store the starting key of each child and the hash of the child node. The hash of the root node and other metadata such as block number, timestamp, and transaction IDs, are included in a data block, which is stored as a leaf of the upper level POS-tree. The keys of the upper level POS-tree are block numbers. The hash of this tree's root is the digest of the entire ledger.
Retrieving a key from a given block number entails getting the data block with the corresponding block number from the upper level POS-tree, then traversing the lower level  POS-tree to locate the data.
When updating a key, new nodes are created at both levels using copy-on-write.

One advantage of this data structure is that it provides efficient current-value proofs, in addition to the
inclusion and append-only proofs. Since each data block represents a snapshot of the database states, the latest values always
appear in the last block, thus current-value proof for different keys can be batched and verified with only one block.
Another advantage is that enables efficient failure recovery. 
In particular, during recovery \systemname re-executes the transactions to update the two-level POS-tree. If failure happens during updating of the lower level tree, all nodes created before failure can be reused, since the POS-tree structure would be the same given the same input (the order of updates does not matter). However, if failure happens during updating of the upper level tree, the nodes created before failure cannot be reused, because the data blocks include varying content like timestamps. To address this, we keep an additional mapping between block sequences and persisted data blocks. If the block sequence already exists, \systemname will only re-execute the updates of the upper-level POS-tree based on the persisted blocks.

\textit{Example.} Consider the two-level POS-tree in Figure~\ref{fig:2lvlpostree}. When $K_9$ is updated with the value $V_9^2$, \systemname  first updates the lower level tree using copy-on-write, creating a new leaf node $b'$ by replacing the $V_9^1$ and $H_\phi$ with $V_9^2$ and $H_b$ respectively. It then updates node $c$ with the hash of node $b'$, creating a new node $c'$. This is done recursively until a new root node $LR2$ is created. Next, \systemname uses the hash of $LR2$ to create node $BLK2$ with block number $B$ and timestamp $t_B$. Finally, it propagates the update towards the root of the upper level tree.

The inclusion proof of $K_9$ at $B-1$ includes the nodes $b$, $c$, $LR1$, $BLK1$, $d$, $UR1$, and $D_1$. The verification is done by recursively computing the hash of the child node and comparing it with what is stored in the parent node, and finally checking that $H(UR1)$ is equal to $D_1$. The current-value proof is generated by computing the inclusion proof based on the last block, e.g. $BLK2$. 
The append-only proof for showing that the ledger corresponding to $D_1$ is a prefix of the ledger corresponding to $D_2$ includes all the common ancestors of $BLK1$ and $BLK1+1$ in the tree whose digest is $D_2$. In Figure~\ref{fig:2lvlpostree}, the proof includes $d'$, $UR2$, and $D_2$. The verification is done by checking that $d$ is the prefix of $d'$, and the path from $d'$ to $D_2$ is correct.

\noindent
{\bf Discussion.}
The ledger storage consists of two main components: the ledger structure for storing transactions, and the index for accessing the states. \systemname has a smaller ledger structure than QLDB and LedgerDB, because its upper POS-tree stores multiple transactions in one block, whereas QLDB and LedgerDB stores one transaction per leaf node of their Merkle trees.  SQL Ledger has a smaller ledger structure than \systemname, because it is based on rows updated within a transaction and transactions committed within a block. However, it uses a hashed chain instead of a Merkle tree, therefore it is less efficient in verification. 
QLDB and SQL Ledger do not protect the index, thus the server can return stale data, or tamper with the indexes without being detected. LedgerDB constructs clue indexes and protects the size of clue indexes using an additional Merkle Patricia Trie. However, the server can still modify the pointers inside the skip lists to point to stale entries. To detect such tampering, the client needs to verify all entries in the skip lists, thus incurring significant costs. In contrast, the lower level POS-tree in \systemname both protects the index and provides efficient access.

\subsubsection{Transaction}
\label{subsubsec:transaction}

\systemname partitions the keys into shards based on their hash values.
When a transaction involves multiple shards, 
\systemname achieves atomicity using 2PC.
Each client is a coordinator. It generates the read set and write set of the transaction, then sends prepare message to the shards. The transaction manager at each shard logs the transaction and responds with a commit or abort based on the concurrency control algorithm. \systemname uses optimistic concurrency control to achieve serializability. In particular, the read set and write set of concurrent transactions are validated to check for the read-write and write-write conflicts. The shard returns ``commit'' if there are no conflicts, and returns ``abort'' otherwise. The client waits for the responses from all shards involved in the transactions, and it resends the messages after a timeout period. If all shards return commits, the client sends the commit messages to the shards, otherwise it sends abort. Each shard then commits or aborts the transaction accordingly, and returns an acknowledgment to the client. 

At each shard, the transaction is processed by the transaction manager as follows. All incoming requests are buffered in the transaction queue, waiting to be assigned to available transaction threads. If the queue is full, the transaction is aborted.  The transaction threads store the prepared transactions and committed data in the shared memory. The persisting thread persists the committed data asynchronously to the ledger storage.

\noindent
\textbf{Asynchronous persistence.} Committing transactions to the ledger incurs large
overheads due to high contention and long execution time.  To address this, \systemname updates the ledger asynchronously. In particular, when receiving the commit message, the transaction manager stores the transaction data in a multi-version ``committed data map'' $\langle \textit{key, ver, val}\rangle$
in memory, and writes to the WAL for durability and recovery. After a
timeout, a background thread persists the data in the map to the
ledger storage. The persisted data is then removed from the committed data map
to keep the memory consumption low. This approach moves the updating of the ledger out
of the critical path, thus reducing transaction latency. The trade-off here is that the users cannot retrieve the proofs for data that has not been persisted to the ledger. We explain the verification process in Section \ref{subsec:verification}. 

\noindent
\textbf{Transaction batching.}
The cost of updating and persisting the authenticate data structures is large, even though they are now out of the critical path of transaction execution. It is because  both levels of the POS-tree need to be updated and written to disk. 
To reduce this cost, \systemname batches multiple committed transactions
before updating the ledger. In particular, it uses an aggressive batching strategy that collects independent data from recently committed transactions into a data block.
All the  blocks created within a time window are appended to the ledger storage.
To form a block, the server selects data from the ``committed data map'' version by version.
For a given data version, it can compute the sequence number of the block at which the data will be committed, by adding the current block sequence with the version sequence in the data map. This estimation is used for deferred verification (explained later in Section \ref{subsec:verification}). 


\noindent
{\bf Discussion. }
Both LedgerDB and SQL Ledger support asynchronous persistence and transaction batching. 
However, \systemname maintains the key-value mapping in memory, and only writes WAL to disk during persistence. Therefore, it incurs a lower commitment cost than the other two systems. The batching in \systemname takes advantage of its index structure to improve performance. In particular, it batches non-overlapping keys from multiple transactions into one block and builds upper level POS-tree on the blocks, which leads to a smaller ledger structure, and consequently lower verification cost, i.e., $O(log B)$. On the other hand, LedgerDB creates a block for each transaction when committing, and batch updates the Merkle tree with multiple blocks periodically. SQL Ledger batches multiple transactions in a Merkle tree and appends a new block created with the Merkle tree root to a hashed chain of blocks. The cost of verification is $O(log T)$ for LedgerDB and $O(log T/B + m)$ for SQL Ledger, where $T$ is the total number of transactions, $T/B$ is the number of transactions for a batch, and $m$ is the number of blocks in the hash-chain scanned. These costs are greater than $O(log B)$.

\subsubsection{Verification}
\label{subsec:verification}

Verifying a transaction requires checking both the read set and the write set. To verify the read set, the client checks that the data is correct and is the latest, 
i.e., current-value proof.
To verify the write set, the client checks that the new ledger is append-only and that the data written to the ledger is correct,
i.e., append-only proof and inclusion proof.
As an example, consider a client holding a stale digest $D_0$ commits a transaction that performs read-modify-write on the key $K_9$. For verification, the client requests four proofs: an append-only proof of current digest $D_1$ from $D_0$, a current-value proof of $K_9$ and $V_9^1$ with respect to $D_1$, an inclusion proof of $K_9$ and $V_9^2$ with respect to the new digest $D_2$, and an append-only proof of $D_2$ from $D_1$.
In \systemname, the verification requires getting proofs from all participating shards.
There is no coordination overhead, because the ledger is immutable with copy-on-write which means verification can run concurrently with other transactions.  

\noindent
{\bf Deferred verification. }
\systemname supports {\em deferred} verification, meaning that transaction verification occurs within a time window, as opposed to immediately. This strategy is suitable for applications that require high performance and can tolerate temporary violations of data integrity.
For these applications, the client gets a {\em promise} from the server containing the
future block sequence number where the data will be committed, transaction ID, current digest, the key and the value. The client can verify the transaction after the block is available by sending a verification request taking the promise as the parameter. The server, on receiving the verification request, will check if the block has been persisted. It generates the inclusion proof and append-only proof if the check passes, and returns the proofs and new digest to the client. The client can then verify the integrity of the data as mentioned above. The two-level POS-tree allows the server to batch proofs for multiple keys (especially when they are packed in the same data block). Furthermore, getting the data and the proof can be done at the same time by traversing the tree, which means proof generation can be done with little cost when fetching the data during transaction processing. This is as opposed to LedgerDB requiring the server to traverse one data structure to retrieve the data, and then another data structure to retrieve the proof. To alleviate the burden of deferred verification, in \systemname, the proof of persisted data is returned immediately during transaction processing, and proof for data to be persisted in future blocks will be generated in deferred verification requests in batches.
Deferred verification in \systemname makes the batching of proof more effective than in LedgerDB and SQL Ledger. This is because the system only needs to access the last block to generate the proof, since the last block covers all current values.

This approach leaves a window of vulnerability during which a malicious
database can tamper with the data, but any misbehavior will be detected once the {\em promised} block number appears in the ledger.
\systemname allows clients to specify customized delay time for verification to find suitable trade-offs between security guarantee and performance according to their needs. 
Particularly, zero delay time means immediate verification. In this case, the transactions are persisted in the ledger synchronously during the commit phase.
This strategy is suitable for applications that cannot afford even a temporary violation of data integrity.

\subsubsection{Auditing}


While the user verification ensures that the user's own transactions are executed
correctly, \systemname ensures the correct execution of the database server
across multiple users. In particular, it relies on a set of auditors, some of
which are honest, to ensure that different users see consistent views of the
database.

Each auditor performs two important tasks. First, it checks that the
server does not fork the history log, by checking that the users receive
digests that correspond to a linear history. It maintains a current
digest $d$ and block number $b$ corresponding to the longest history that it
has seen so far. When it receives a digest $d'$ from a user, it
asks the server for an append-only proof showing that $d$ and $d'$ belong to a
linear history.

Second, the auditor re-executes the transactions to ensure that the current
database states are correct. This is necessary to prevent the server from
arbitrarily adding unauthorized transactions that tamper with the states. It
also defends against undetected tampering when some users do not perform
verification (because they are offline, or due to resource constraints). The 
auditor starts with the same initial states as the initial states at the 
server. For each digest $d$ and corresponding block number $b$, the auditor
requests the signed transactions that are included in the block, and the proof
of the block and of
the transactions.  It then verifies the signatures on the transactions,
executes them on its local states, computes the new digest, and verifies it
against $d$.

When the auditor receives a digest corresponding to a block number $b'$ 
which is larger than the current block number $b$, it first requests and verifies the
append-only proof from the server. Next, for each block between $b$ and $b'$,
it requests the transactions and verifies that the states are updated
correctly. After that, it updates the current digest and block number to $d'$ 
and $b'$ respectively. Finally, after a pre-defined interval, the auditor
broadcasts its current digest and block number to other auditors.

\remove{
While the client verification ensures the clients' own transactions are executed correctly, it is the auditors' job to ensure that transactions issued by other clients and the entire system are in correct states. \systemname relies on a group of auditors, which gossip among themselves to ensure global consistency. There are two tasks of auditors. First, the auditors synchronize with the servers and among each other to maintain a consistent history of digests. Second, they replay the transaction log to ensure the state of the database is correct. To achieve the first task, the auditor broadcasts the latest block number and digests obtained from the server to the group of auditors. The recipients which possess larger block number will check whether the digests exist in its digest history. To achieve the second task, the auditor requests the server for proof step-by-step for each digest. The servers return to the auditor the transaction entries included with clients' signature, the tip block data 
for that digest, and the proof of the block. The auditor validates the signature with the public key received from the clients on their initialization and replays the transaction entries. The root hash is compared with the data hash in the block data. Lastly, the auditor verifies the block data by recalculating the digest with the proof.
}

\subsubsection{Failure Recovery}
\label{subsec:ft}

\systemname supports transaction recovery after a node crashes and reboots.
In particular, if a node fails before the commit phase, the client aborts the transaction after a timeout.
Otherwise, the client proceeds to commit the transaction.  When the failed node
recovers, it queries the client for the status of transactions, then decides to
whether abort or commit.  It then checks the WAL for updates that have not been
persisted to the ledger storage, and updates the latter accordingly. If the
client fails, the nodes have to wait for it to recover, because the 2PC
protocol is blocking. We note that this can be mitigated by replacing 2PC with
a non-blocking atomic commitment protocol.
For example, three-phase commit (3PC) uses an extra phase, allowing participants to communicate among themselves. Another example is non-blocking 2PC~\cite{easycommit} that requires participants to forward the vote decisions to every other node. Paxos commit~\cite{paxoscommit} is also non-blocking, in which additional nodes called {\em acceptors} ensure that the votes are not lost in case of failure. However, these protocols are more complex and incur higher network overheads than 2PC. Integrating them to \systemname is left as future work.

\systemname tolerates permanent node failures by
replicating the nodes using a crash-fault tolerant protocol, namely Raft. To ensure consistent ledgers across the replicas, \systemname uses a fixed batch size when creating the blocks during the persistence phase. A timeout is set in case the number of upcoming transactions is insufficient to build a block.
When it is expired, a dummy transaction is replicated to all replicas to enforce the block creation. We evaluate the performance impact of node crashes on both schemes in section \ref{sec:exp}.

\subsection{Analysis}

\subsubsection{Cost analysis}
Similar to other verifiable databases, \systemname incurs additional costs to
maintain the authenticated data structure and to generate verification proofs
compared to conventional databases. We now analyze the asymptotic computational
costs of the main operations in \systemname.

\noindent
{\bf Persistence.}
The persistence phase updates the committed data to the two-level POS-tree. The
cost of this phase is bounded by the height of the tree, which is the height
of upper level plus that of the lower level, i.e., $O(\log{B} + \log{m})$, where B
is the number of blocks and m is the number of distinct keys. In contrast, LedgerDB
needs to update both the bAMT and ccMPT, which is $O(\log{N} +
\log{m})$, where N is the total number of transactions.  We note that due to
batching, $B$ is much smaller than $N$.  QLDB also updates the Merkle
tree over the transactions, thus its cost is $O(\log{N})$.

\noindent
{\bf Inclusion proof.}
To generate an inclusion proof, \systemname traverses the two-level POS-tree
to get the nodes on the path from the leaf to the root. Hence, the
cost is $O(\log{B} + \log{m})$. For LedgerDB and QLDB, the proof
includes the Merkle proof for a transaction, and the transaction content. The
cost is $O(\log{N})$, since the transaction content is small compared
to the number of transactions.

\noindent
{\bf Current-value proof.}
In \systemname, the lower-level POS-tree captures the entire states, therefore
the latest value always appears in the right-most block.  The current-value proof
is a special case of inclusion proof that includes the right-most block. In other
words, the cost is $O(\log{B} + \log{m})$.  In contrast, LedgerDB and QLDB do
not have protection over indexes.  Therefore, their current-value proofs
require scanning from one transaction to the latest transaction to
check for any new updates on the key. The cost of this is $O(N)$.

\noindent
{\bf Append-only proof.}
The append-only proof checks if the two digests belong to a linear history.
Such a proof contains the nodes created between one digest and another. The cost is $O(\log{B})$ for \systemname,
and $O(\log{N})$ for LedgerDB and QLDB.

\subsubsection{Security analysis}
\systemname is a verifiable ledger database system since it supports the four
operations described in Section~\ref{sec:vldb}. We now sketch the proof that
\systemname is secure, which involves showing that it satisfies both integrity
and append-only property.

For integrity, we first consider Get operation that returns the latest value of
a given key (the other Get variants are similar) at a given $\digest$. The user
checks that the returned proof $\pi$ is a valid inclusion proof corresponding
to the latest value of the key in the POS-tree whose root is $\digest$. Since
POS-tree is a Merkle tree, integrity holds because a proof to a different value
will not correspond to the Merkle path to the latest value, which causes the
verification to fail. Next, consider the Put operation that updates  a key. The user verifies that the new value is included as the latest value
of the key in the updated digest. By the property of the POS-tree, it is not
possible to change the result (e.g., by updating a different key or  updating the given key with a
different value) without causing the verification to fail.

For append-only, the auditor keeps track of the latest digest $\digest_{S,H}$
corresponding to the history $H$. When it receives a digest value
$\digest_{S',H'}$ from a user, it asks the server to generate an append-only
proof $\pi \leftarrow \ProveAppend(\digest_{S',H'}, \digest_{S,H})$. Since our
POS-tree is a Merkle tree whose upper level grows in the append-only fashion,
the server cannot generate a valid $\pi$ if $H'$ is not a prefix of $H$ (assuming $|H'| < |H|$). Therefore, the append-only property is achieved.

In \systemname, each individual user has a local view of the latest digest
$\digest_{S_l,H_l}$ from the server. Because of deferred verification, the user
sends $\digest_{S_l,H_l}$ together with the server's promise during verification.
When the latest digest at the server $\digest_{S_g,H_g}$ corresponds to a
history log $H_g$ such that $|H_g| > |H_l|$, the server also generates and
includes the proof $\pi \leftarrow \ProveAppend(\digest_{S_l,H_l},
\digest_{S_g,H_g})$ in the response to the user.  This way, the user can detect
any {\em local} forks in its view of the database. After an interval, the user
sends its latest digest to the auditor, which uses it to detect {\em
global} forks.


\section{Benchmark}
\label{sec:benchmark}
Even though the verifiable databases expose database-like interface to applications, the existing database benchmarks do not contain verification workloads. To fairly compare \systemname with other systems, we extend YCSB and TPC-C by including verification workloads.

\subsection{YCSB}
The existing YCSB workloads include simple put and get operations.  We add three more operations, called {\tt
VerifiedPut}, {\tt VerifiedGetLatest}, and {\tt VerifiedGetHistory}, and a new parameter {\tt delay}. These
operations return integrity proofs that can be verified by the user. The {\tt delay} parameter allows for
deferred verification, that is, the database generates the proofs only after the specified duration. When
set to $0$, the operations return the proof immediately. When greater than $0$, the database can improve
its performance by batching multiple operations in the same proof.
\begin{itemize}[leftmargin=*]
\item {\tt VerifiedPut(k,v,delay)}: returns a promise. The user then invokes {\tt GetProof(promise)} after
  {\tt delay} seconds to retrieve the proof.

\item {\tt VerifiedGetLatest(k,fromDigest,delay)}: returns the latest value of $k$. The user only sees the
  history up to {\tt headDigest}, which may be far behind the latest history. For example, the user last
  interacts with the database, the latter's history digest is {\tt fromDigest}. After a while, the history is
  updated to another digest {\tt latestDigest}. This query allows the user to specify the last seen history.
  The integrity proof of this includes an append-only proof showing a linear history from {\tt fromDigest} to
  {\tt latestDigest}.

\item {\tt VerifiedGetHistory(k,atDigest,fromDigest)}: returns the value when the database history is at {\tt
  atDigest}. {\tt fromDigest} is the last history that user sees. The integrity proof for this query includes
  an append-only proof from {\tt fromDigest} to {\tt atDigest}.
\end{itemize}
Based on these operations, we add two new  workloads to YCSB. First, 
{\em Workload-X} consists of  50\% VerifiedPut, 50\% VerifiedGetLatest, with 100ms delay.
Second, {\em Workload-Y} consists of 20\% VerifiedPut, 40\% VerifiedGetLatest, 40\% VerifiedGetHistory, with 100ms delay.

\subsection{TPC-C}
\label{subsec:tpccbench}
We extend all five types of transactions in TPC-C to verified versions. Similar to YCSB, each new
transaction has a {\tt delay} parameter for specifying deferred verification. When {\tt delay > 0}, each
transaction returns a promise which is later used to request the integrity proof. In addition to the five new
transactions, we add a new one called {\tt VerifiedWarehouseBalance}, which retrieves the last $10$ versions
of {\tt w\_ytd}. This transaction is possible with verifiable ledger databases because they maintain all
historical versions of the data.



\section{Evaluation}
\label{sec:exp}

\subsection{Baselines}
\label{subsec:baseline}

We compare \systemname against four state-of-the-art verifiable ledgers, namely QLDB, LedgerDB, SQL Ledger and Trillian. We do not compare against blockchains due to the different threat models. QLDB, LedgerDB, and SQL Ledger are not open-sourced, thus we implement them based on the documentation available online, or based on the details in the papers.
We denote them by QLDB$^\ast$, LedgerDB$^\ast$, and SQL Ledger$^\ast$ respectively.

To facilitate \textit{fair performance comparison},  we
implement QLDB$^\ast$, LedgerDB$^\ast$, SQL Ledger$^\ast$ and \systemname on top of the
same distributed layer, which removes the impact of communication protocols and related implementation details on the overall performance gaps. In particular, all four systems partition their
data over multiple nodes, and they use the same 2PC implementation for
distributed transactions. All systems are implemented in C++, using libevent v2.1.12 and Protobuf v3.19.3 for network communication and serialization respectively. We use BLAKE2b as the cryptographic hash function. 




\textit{QLDB$^\ast$.}
\label{subsection:emulated_system}
We implement a version of QLDB based on the available documentation.
The system consists of ledger storage and index storage as shown in Figure~\ref{fig:qldb}.
The former maintains the transaction log (the WAL) and a Merkle tree is built on top of it.
The latter maintains a B$^+$-tree and data materialized from the ledger.
When committing a new transaction, the system appends a new log entry
containing the type, parameters and other metadata of the transaction, and updates
the Merkle tree.  After that, the transaction is considered committed and the
status is returned to the client.  A background thread executes the
transaction and updates the B$^+$-tree with the transaction data and metadata.
We implement the inclusion proof, append-only proof, and current-value proof as
described in Section~\ref{subsec:review}. We do not implement the SQL layer,
which is complex and only negatively impacts the overall performance of OLTP workloads.


\textit{LedgerDB$^\ast$.}
We implement a version of LedgerDB based on the descriptions in \cite{ledgerdb}.
The system consists of a transaction log, a bAMT, clue indexes, and a ccMPT.
We implement bAMT by
forcing the update API to take as input a batch of transaction journals.  The
nodes in bAMT are immutable, that is, each modification results in a new node.  We
implement the clue index by building a skip-list for each clue, and keeping
the mapping of each clue to its skip list's head in memory. To enable queries on keys, we
create one clue for each key. As the result, a  transaction may update multiple skip lists.
The ccMPT is constructed to protect the integrity of the clue index, in which the key
is the clue, and the value is the number of leaf nodes in the corresponding
skip-list.  When committing a new transaction, the system creates a transaction
journal containing the type and parameters of the transaction, updates the clue
index with it, and returns the status to the client.  A background thread
periodically updates the bAMT with the committed journals, and updates the ccMPT. 
The root hash of ccMPT and bAMT are stored in a hash
chain. We do not implement the timeserver authority (TSA), which is used for
security in LedgerDB, as its complexity adds extra overhead to the system.

\textit{SQL Ledger$^\ast$.} We implement SQL Ledger based on the original paper~\cite{sqlledger}. To fairly compare with other systems, we omit the SQL layer and only implement the key-value data model. We use B$^+$-trees for the indices. When committing data, we first append the log entry to the WAL, then update the current and history B$^+$-tree. Next, we create a list of data entries for all modified data within the transaction and append them to an in-memory queue. A background thread builds a Merkle tree and creates a transaction entry using the Merkle tree root hash, transaction ID, and timestamp for each data entry list stored in the queue. After that, the Merkle tree for transactions is constructed based on the transaction entries. Finally, a block containing the root hash of  the Merkle tree for transactions, block sequence, and previous block hash is appended to a hash chain.

\textit{Trillian.} We use an implementation of verifiable log-based maps provided by \cite{wave, trillian}.
The system exposes a key-value interface, and consists of two transparency logs and one map. It stores the map of
all the keys in a sparse Merkle tree, and asynchronously updates the maps with the new operations in batches. 
We use the default
configurations provided in~\cite{wave}, and use the throughput of the root log as the overall throughput.

\remove{When a new key is added, the operation is appended to the {\em operation log}. The system stores the map of
all the keys in a sparse Merkle tree, and asynchronously updates the maps with the new operations in batches.
Each update results in a new Merkle root, which is then appended to the {\em root log}. An auditor process
executes the update operations and checks that the new Merkle roots are computed correctly. Every update
operation returns a {\em promise} indicating the future version where the update will be reflected in the map.
Each get operation returns the inclusion proof, and append-only proof of the root log. We use the default
configurations provided in~\cite{wave}, and use the throughput of the root log as the overall throughput.}

\subsection{Experiment Setup and Results Summary}
All experiments are conducted on 32 machines with Ubuntu 20.04, which are equipped with 10x2 Intel Xeon CPU W-1290P processor (3.7GHz) and 125GB RAM. The machines are on the same rack and connected by 1Gbps network.
For each experiment, we collect the measurements after a warm-up of two minutes during which the systems are stable. The results show that \systemname consistently outperforms QLDB$^\ast$, LedgerDB$^\ast$, and SQL Ledger$^\ast$ across all workloads. In particular, compared to LedgerDB$^\ast$ it achieves up to $1.7\times$ higher throughput on the YCSB workload, $1.3\times$ higher throughput for the TPC-C workload, and $1.6\times$ higher throughput for the verification workload. These improvements are due to \systemname's efficient authenticated data structure, deferred verification, and effective batching.

We note that our results are different from what is reported in~\cite{ledgerdb}. In particular, the absolute and relative performances of different systems are not the same, for which we attribute to four reasons. First, the Amazon QLDB service has low performance due to its many limitations such as the maximum transaction size. Moreover, it runs on a serverless platform that is out of the user's control, making it difficult for fair comparisons. Our emulated implementation of QLDB removes these limitations and achieves much higher throughput than the original Amazon QLDB service. Second, \cite{ledgerdb} lacks sufficient details regarding the experiments. For example, it does not say how it is compared against QLDB (whether the authors used the QLDB service from Amazon, or implemented an emulated one). Third, since the technique details of QLDB are not public, it is not clear from ~\cite{ledgerdb} how much the performance gap is due to the differences in hardware, low-level communication protocol, or serialization. Finally, it is not possible to reproduce the results in \cite{ledgerdb} due to the experiment artifacts being unavailable. 






\subsection{Micro-Benchmarks}

\begin{figure*}[t]
    \centering
    \begin{minipage}{\linewidth}
        \centering
        \subfigure[Varying transaction sizes]{
            \centering
            \includegraphics[height=2.9cm]{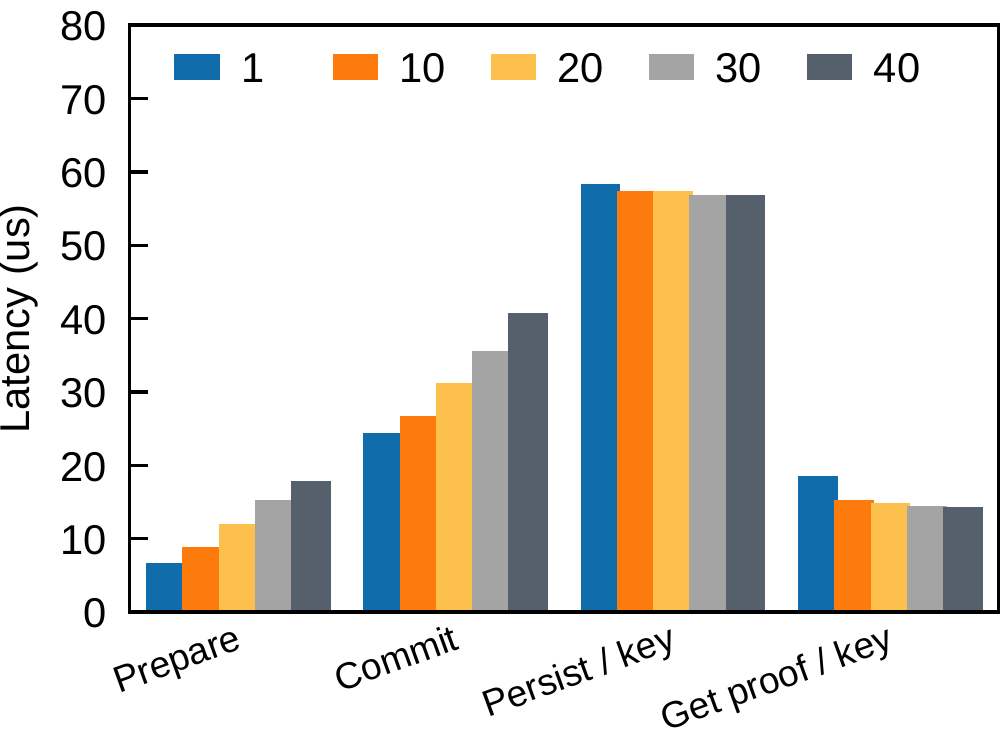}
            \label{fig:exp:bd_tlen}
        }
        \subfigure[Varying workloads]{
            \centering
            \includegraphics[height=2.9cm]{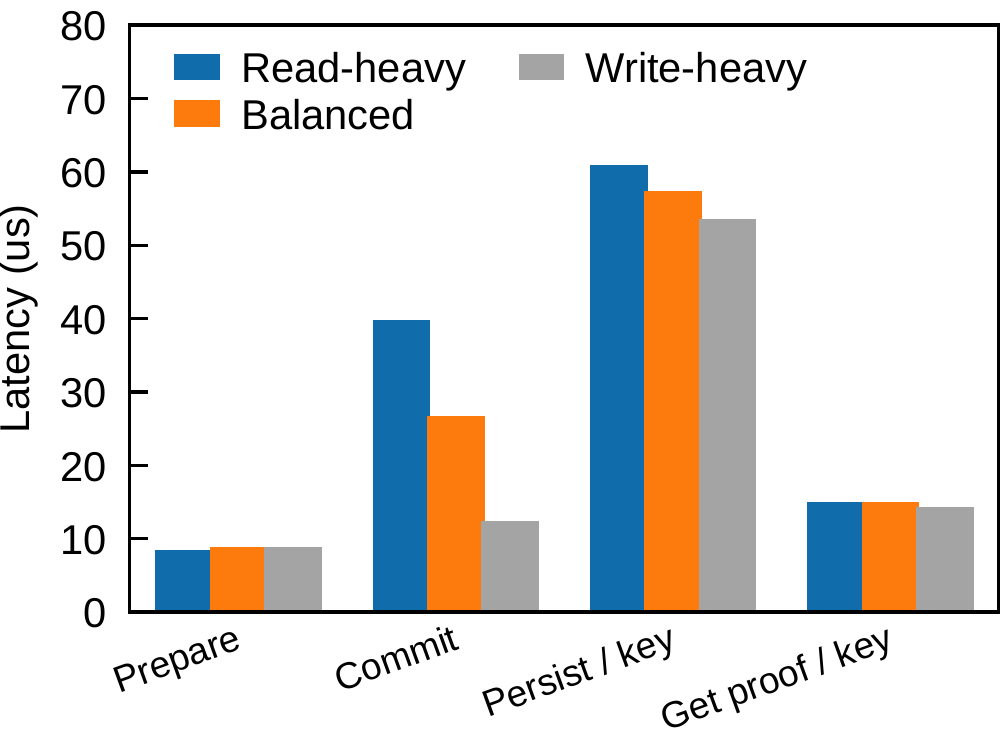}
            \label{fig:exp:bd_workload}
        }
        \subfigure[Varying number of nodes]{
            \centering
            \includegraphics[height=2.9cm]{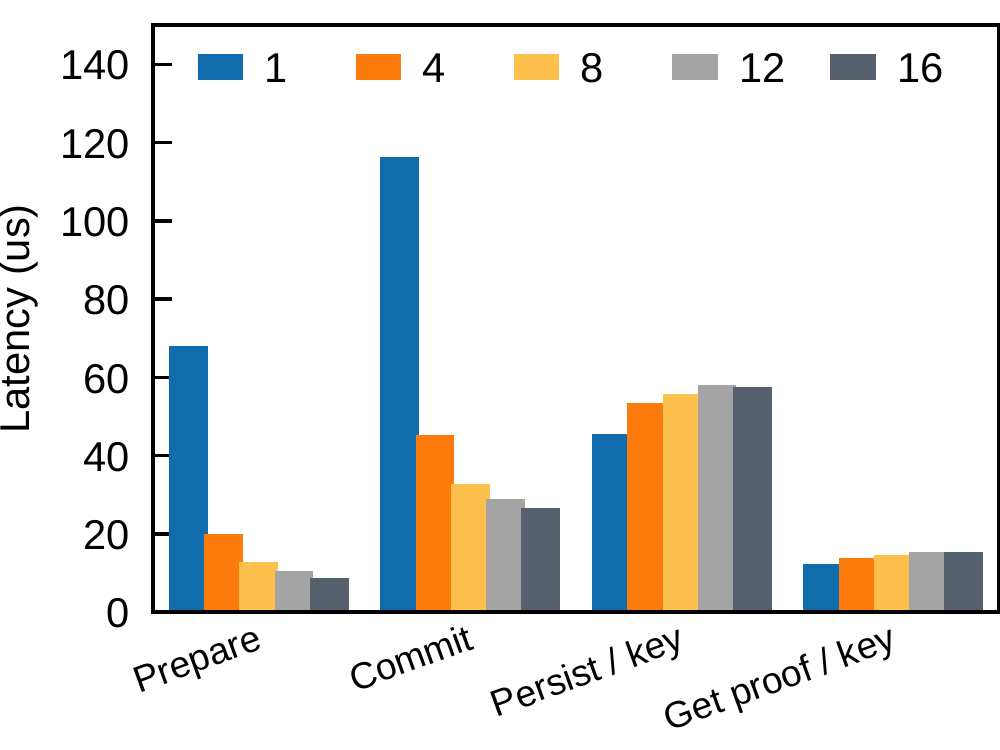}
            \label{fig:exp:bd_server}
        }
        \subfigure[Varying persist interval]{
            \centering
            \includegraphics[height=2.9cm]{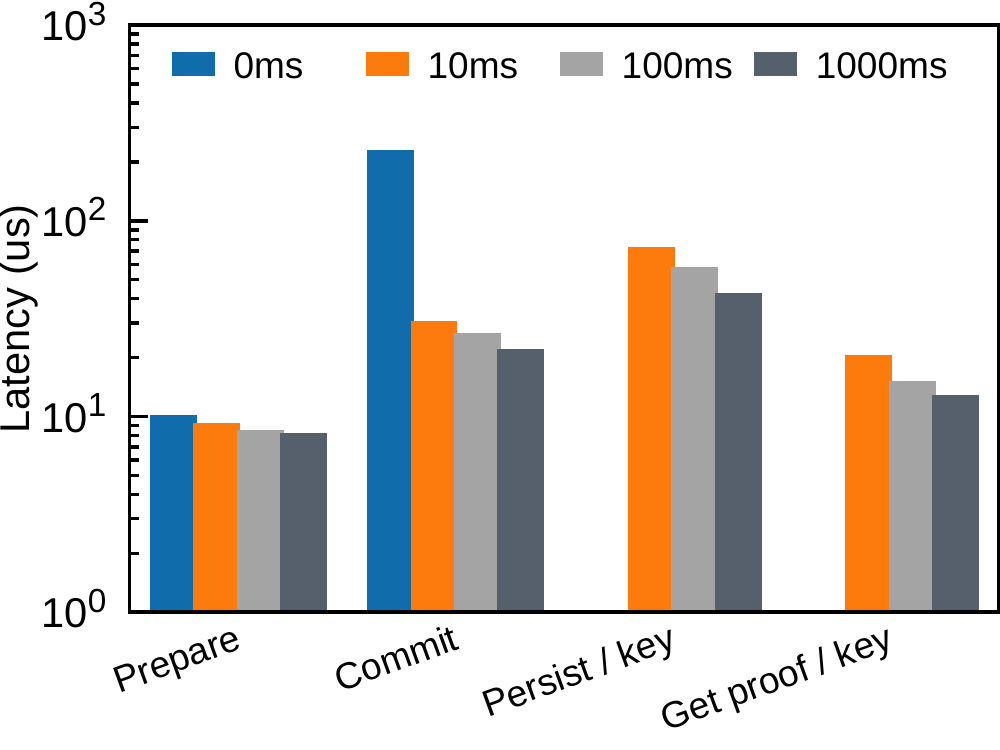}
            \label{fig:exp:bd_delay}
        }
        \vspace{-3mm}
        \caption{Latency breakdown at the server.}
        \label{fig:exp:breakdown}
   \end{minipage}
\vspace{-2mm}
\end{figure*}

\begin{figure*}[t]
    \centering
    \begin{minipage}{0.49\linewidth}
        \centering
        \subfigure[Verification latency at the client]{
            \centering
            \includegraphics[height=2.9cm]{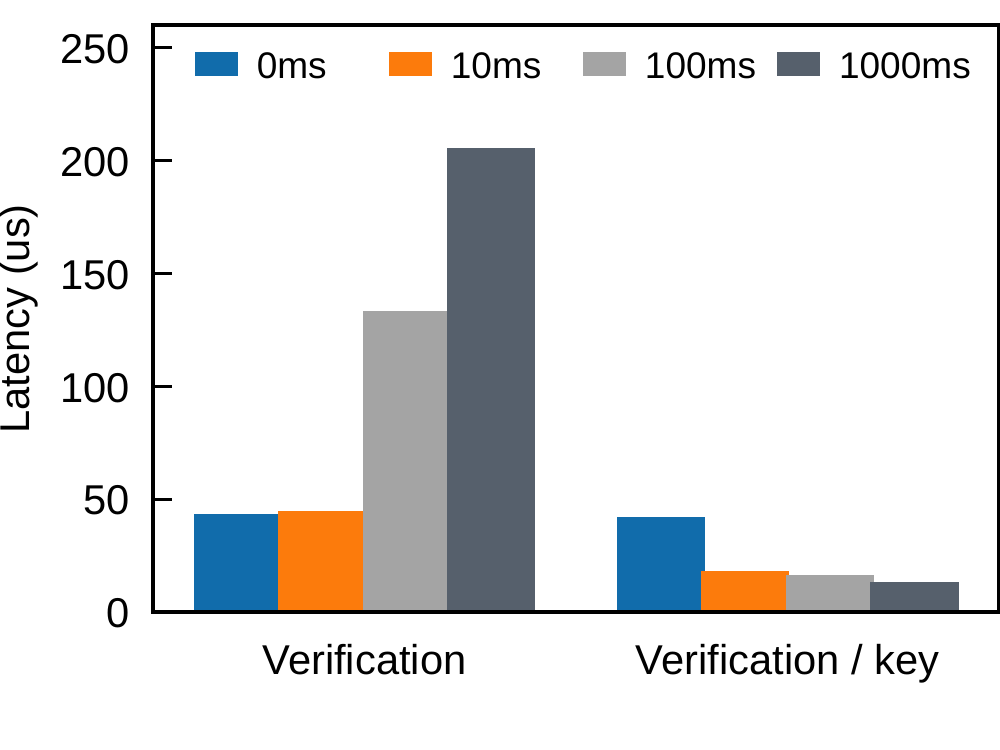}
            \label{fig:exp:bd_delay_per_key}
        }
        \subfigure[Proof size]{
            \centering
            \includegraphics[height=2.9cm]{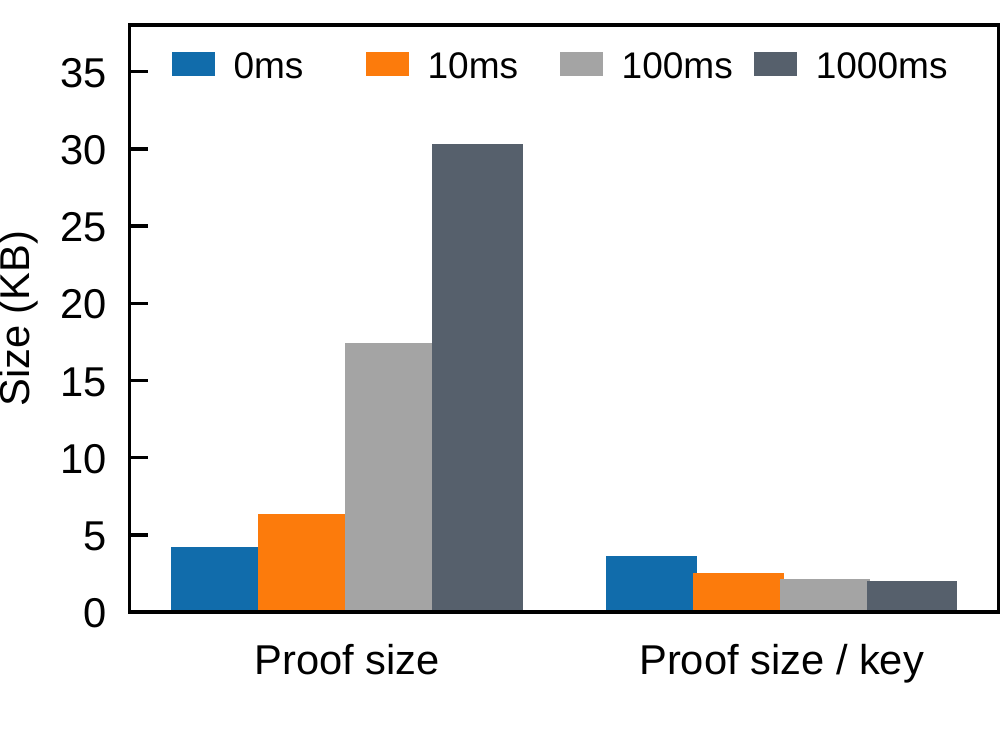}
            \label{fig:exp:bd_delay_proofsize}
        }
        \vspace{-3mm}
        \caption{Impact of delay time at the client.}
        \label{fig:exp:delay}
    \end{minipage}
    \begin{minipage}{0.49\linewidth}
        \subfigure[Varying persist interval]{
            \centering
            \includegraphics[height=2.9cm]{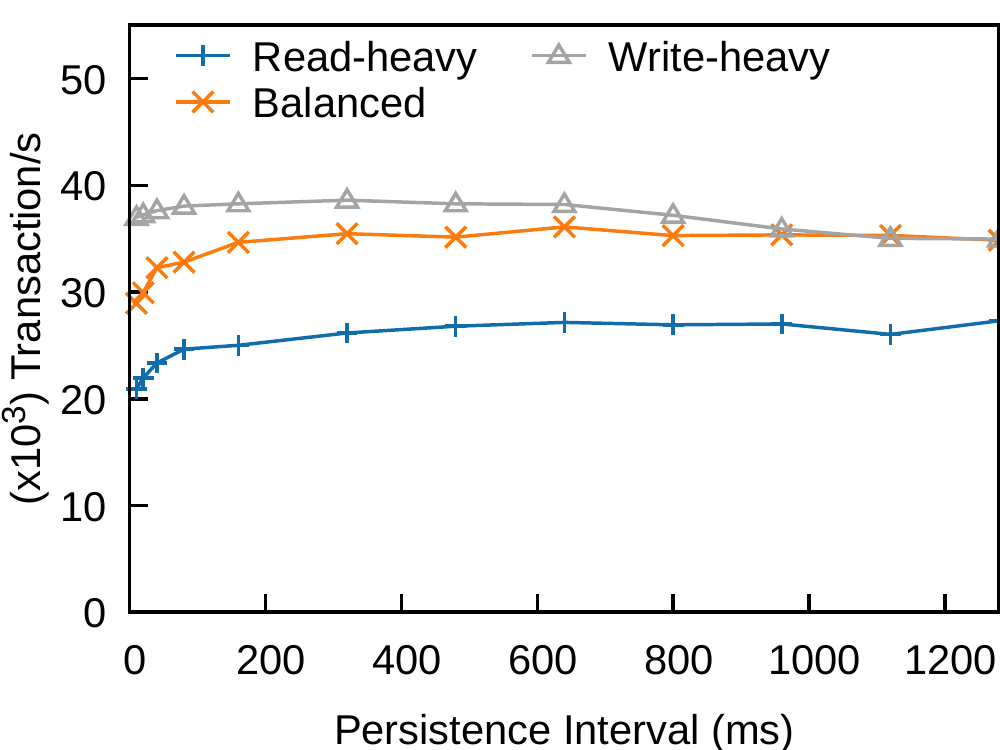}
            \label{fig:exp:delay_persist_tps}
        }
        \subfigure[Varying verification delay]{
            \centering
            \includegraphics[height=2.9cm]{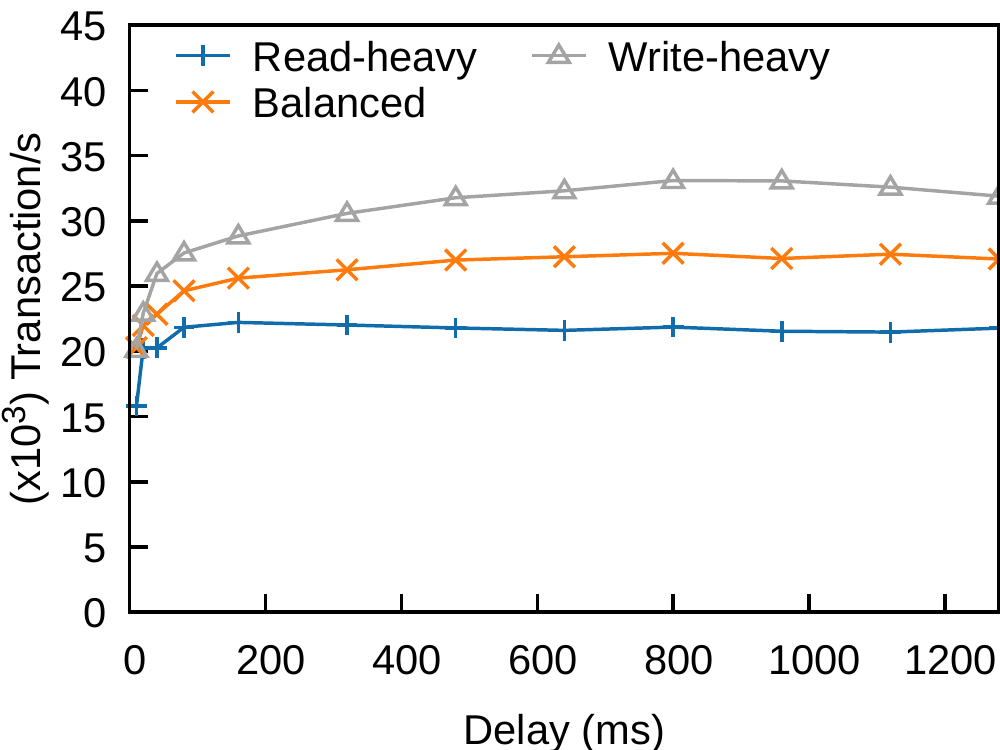}
            \label{fig:exp:delay_verify_tps}
        }
        \vspace{-3mm}
        \caption{Impact of delay time on the overall performance.}
   \end{minipage}
\vspace{-2mm}
\end{figure*}




In this section, we evaluate the cost at the server in terms of  execution time and storage consumption. We extend the vanilla YCSB benchmark to support transactions, by batching every 10 operations as a transaction.  We characterize the workloads as read-heavy (8 reads and 2 writes), balanced (5 reads and 5 writes), and write-heavy(2 reads and 8 writes).

\subsubsection{Cost breakdown of \systemname} We break down the server cost into four phases as described in Section~\ref{subsubsec:transaction}: prepare, commit, persist, and get-proof.
For the last two phases, we report the average latency per key, because these phases' costs depend on the number of records in the batch.

Figure~\ref{fig:exp:bd_tlen} shows the latency of different phases with varying numbers of operations per transaction (or transaction sizes).
We observe that the latency of prepare and commit phase increases as the transactions become larger, which is due to more expensive conflict checking and data operation.
Figure \ref{fig:exp:bd_workload} shows the latency under different workloads.
The latency of the prepare phase increases slightly as the workload move from read-heavy to write-heavy
because a larger write set leads to more write-write and write-read conflict checking.
In contrast, the commit latency of read-heavy workload is much higher than that of write-heavy workload, since read operations are more expensive than the write operations in \systemname as explained in Section \ref{subsec:design:our}. Figure \ref{fig:exp:bd_server} shows the latency breakdown for varying number of nodes. The latency of the prepare and commit phase decrease as the number of nodes increases, because having more shards means fewer keys to process per node.
{Figure \ref{fig:exp:bd_delay} shows the impact of increasing the persist interval. It can be seen that with a longer interval, the persist phase is invoked less frequently, which reduces contention with other phases. As a result, the latency of prepare and commit phases decrease.
The persist batch size increases with longer persist intervals, larger transaction sizes, higher write ratio, or with fewer nodes because they lead to more data committed per node. A large persist batch size results in larger data blocks created in the ledger, which in turn increase the batch size of get-proof phases. The results in Figure~\ref{fig:exp:breakdown} show that persist and get-proof costs decrease as the persist batch size increases, demonstrating the effectiveness of batching.
}


We quantify the cost at the client in terms of verification latency and the proof size (which is proportional to the network cost) as shown in Figure~\ref{fig:exp:delay}. We vary the verification delay to show the impact on the costs. The client batches more keys for verification when the delay time is higher, which results in larger proofs as shown in Figure \ref{fig:exp:bd_delay_proofsize}, and therefore increases the verification latency~\ref{fig:exp:bd_delay_per_key}. We note that the cost per key decreases with higher delay, demonstrating that batching is effective.

We evaluate the impact of the persistence interval on the overall performance by fixing
the client verification delay to 1280ms, while varying the persistence
interval from 10ms to 1280ms.  Figure~\ref{fig:exp:delay_persist_tps} shows the performance for read-heavy, balanced,
and write-heavy workloads. 
{It can be seen that longer intervals lead to higher throughputs
for all workloads except for write-heavy workloads. This is because less frequent updates of
the core data structure helps reduce contention and increase the effect of
batching. For write-heavy workload, however, a long interval  causes the update
of the core data structure to block transaction execution for longer, which increases the abort rate. In particular, we observe that the abort rate
increases to $21.6\%$ at interval of 1280ms for write-heavy workloads, while it remains $1.5\%$ and $3.5\%$
for read-heavy and balanced workloads. Next, we evaluate the impact of verification delay by
fixing the persistence interval to 10 ms and varying the delay from 10 ms to
1280 ms. The results are shown in Figure \ref{fig:exp:delay_verify_tps},
in which the throughput increases with larger delays due to proof batching. 
However, the throughput drops after the peak at 800ms. This is because the batched proof becomes too large that the network cost becomes significant.}

\subsubsection{Cost breakdown versus other baselines}
We compare the latency breakdown of \systemname with that of three other baselines. We do not compare against Trillian because it does not support transactions. 

\begin{figure*}[t]
    \centering
    \begin{minipage}{\linewidth}
        \centering
        \subfigure[Verification time]{
            \centering
            \includegraphics[height=2.9cm]{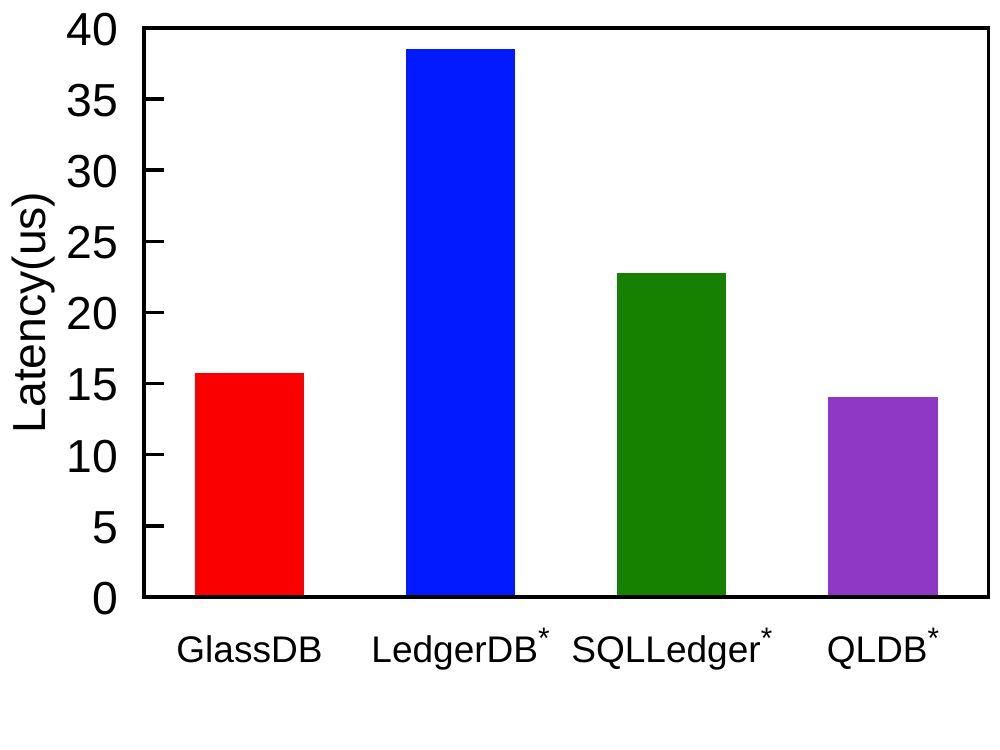}
            \label{fig:exp:proof_latency}
        }
        \subfigure[Proof size per key]{
            \centering
            \includegraphics[height=2.9cm]{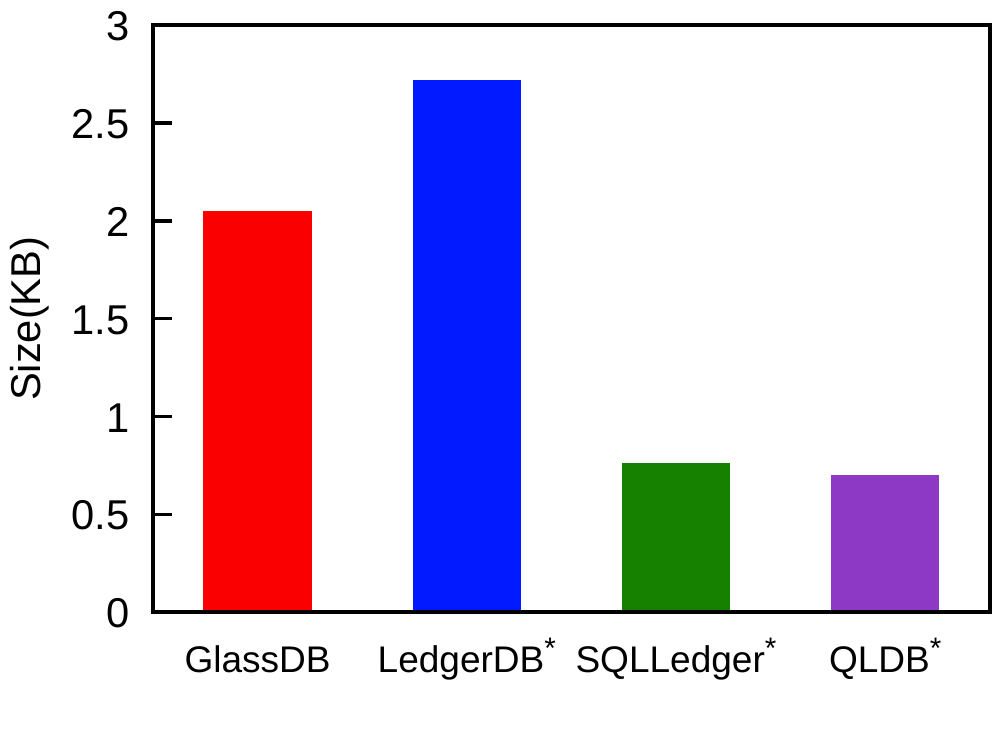}
            \label{fig:exp:proof_size}
        }
        \subfigure[Breakdown latency]{
            \centering
            \includegraphics[height=2.9cm]{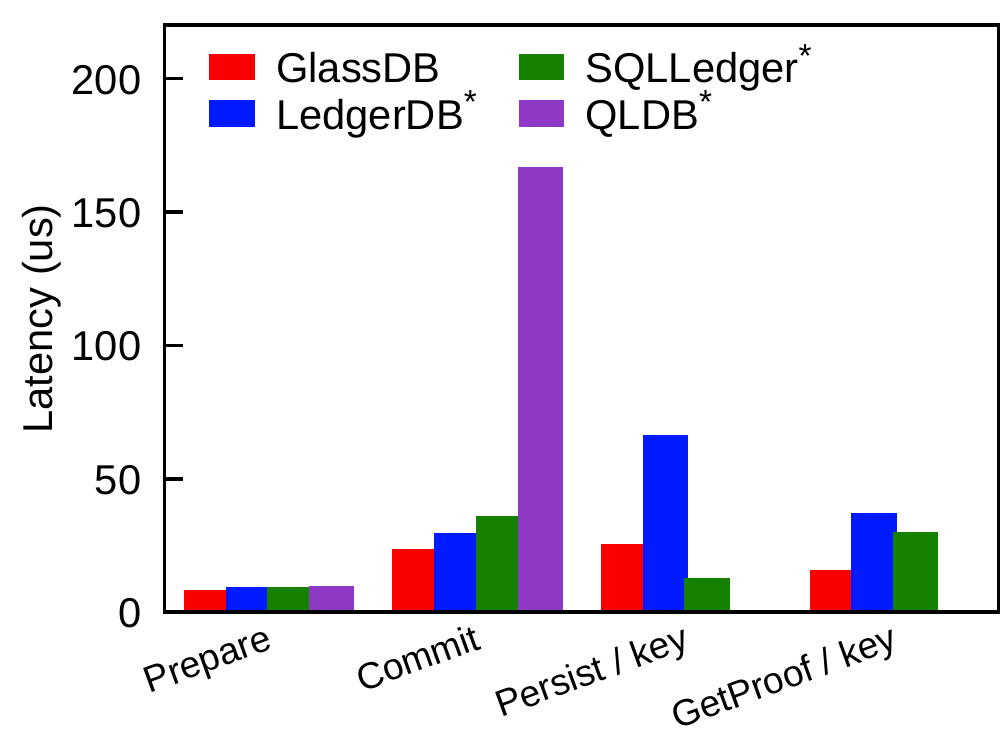}
            \label{fig:exp:bd_comp}
        }
        \subfigure[Storage]{
            \centering
            \includegraphics[height=2.9cm]{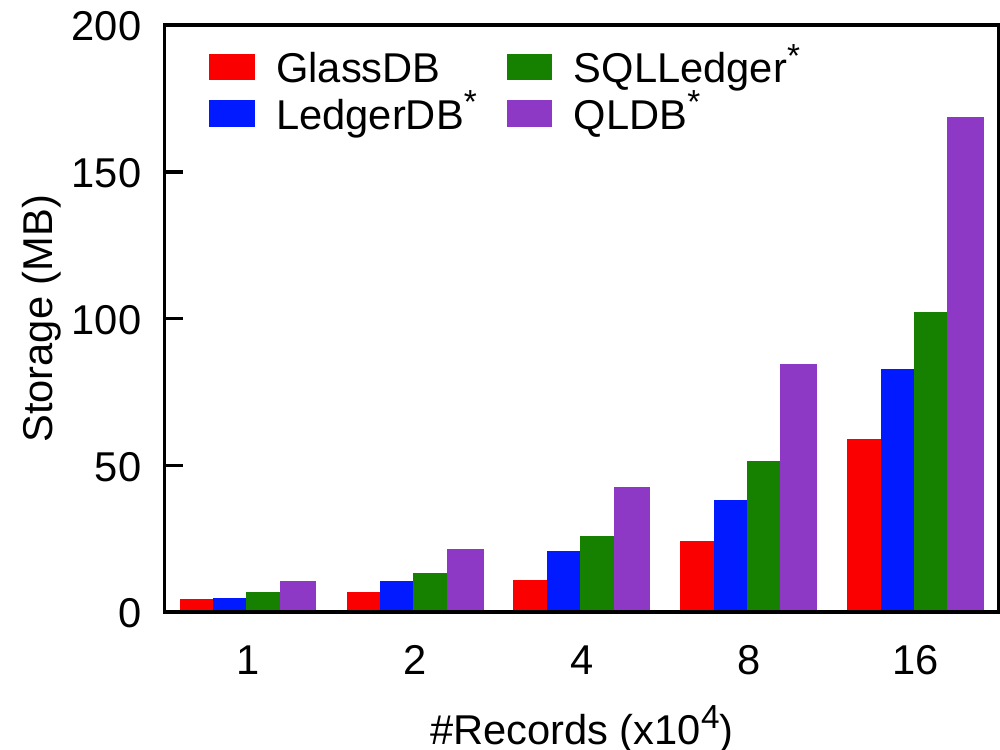}
            \label{fig:exp:storage}
        }
        \vspace{-3mm}
        \caption{Server and client cost versus other baselines.}
   \end{minipage}
\vspace{-2mm}
\end{figure*}

Figure~\ref{fig:exp:proof_latency} and Figure~\ref{fig:exp:proof_size} compare the verification latency and per-key proof size of different systems. We measure the proof size per key because each proof is for the entire block containing multiple keys.
The per-key proof size of QLDB$^\ast$ and SQL Ledger$^\ast$ are smaller than the rest of the systems since they do not have proofs for the  indexes.
QLDB$^\ast$ has the smallest per-key proof size of 0.69KB, and therefore lowest verification time, due to its small Merkle tree. The average Merkle tree height for QLDB$^\ast$ is 17, which is higher than that of SQL Ledger$^\ast$ (height of 12). However, the proof of the latter includes additional hashes of blocks between the target block and the latest block. Therefore, its per-key proof size is slighter bigger, i.e., 0.75KB. \systemname has the smallest tree heights, of 5 and 7 on average for the upper level and lower level POS-tree respectively. However, each node of POS-tree is 4$\times$ larger than that of a Merkle tree, therefore its proof is large, i.e. 2.1KB. Overall, \systemname has a comparable verification time as QLDB$^\ast$, and outperforms SQLLedger$^\ast$. LedgerDB$^\ast$ has the largest tree heights, which are 17 and 19 for bAMT and ccMPT respectively. Therefore, it has the largest per-key proof size and verification time.

Figure \ref{fig:exp:bd_comp} shows that
\systemname, LedgerDB$^\ast$, and SQL Ledger$^\ast$ have lower latency than QLDB$^\ast$ in most phases. The commit latency of QLDB$^\ast$ is especially high because it includes the cost of persisting the authenticated data structure, which explains why Figure~\ref{fig:exp:bd_comp} does not show the cost of the persist phase for QLDB$^\ast$. In contrast, \systemname, LedgerDB$^\ast$, and SQL Ledger$^\ast$ persist the authenticated data structures asynchronously, therefore they have lower latency. 
\systemname has the lowest commit latency because it only persists the write-ahead logs and updates an in-memory map when the transaction commits.
Both LedgerDB$^\ast$ and SQL Ledger$^\ast$ update the index structures during commit. The skip list update in LedgerDB$^\ast$ incurs less overhead than updating the current and history indexes in SQL Ledger$^\ast$. \systemname has lower latency than LedgerDB$^\ast$ in the persist phases because the size of the data committed is smaller. SQL Ledger$^\ast$ commits the least amount of data, therefore it has the lowest persist cost.
\systemname has the lowest per-key latency in the get-proof phase, due to the effective proof batching that reduces the overhead of generating the proof. The batch of get-proof operation for \systemname is 8 on average, while it is 2 for the others.


\vspace{-3mm}
\subsubsection{Storage consumption}
We measure the storage cost of \systemname and compare them against the three other baselines.
Figure \ref{fig:exp:storage} shows that \systemname is the most space-efficient due to the smaller ledger structure and effective batching technique. QLDB$^\ast$ has the highest storage consumption because it updates the ledger for every operation.

\begin{figure*}
    \centering
    \begin{minipage}{0.24\linewidth}
        \centering
        \includegraphics[height=2.9cm]{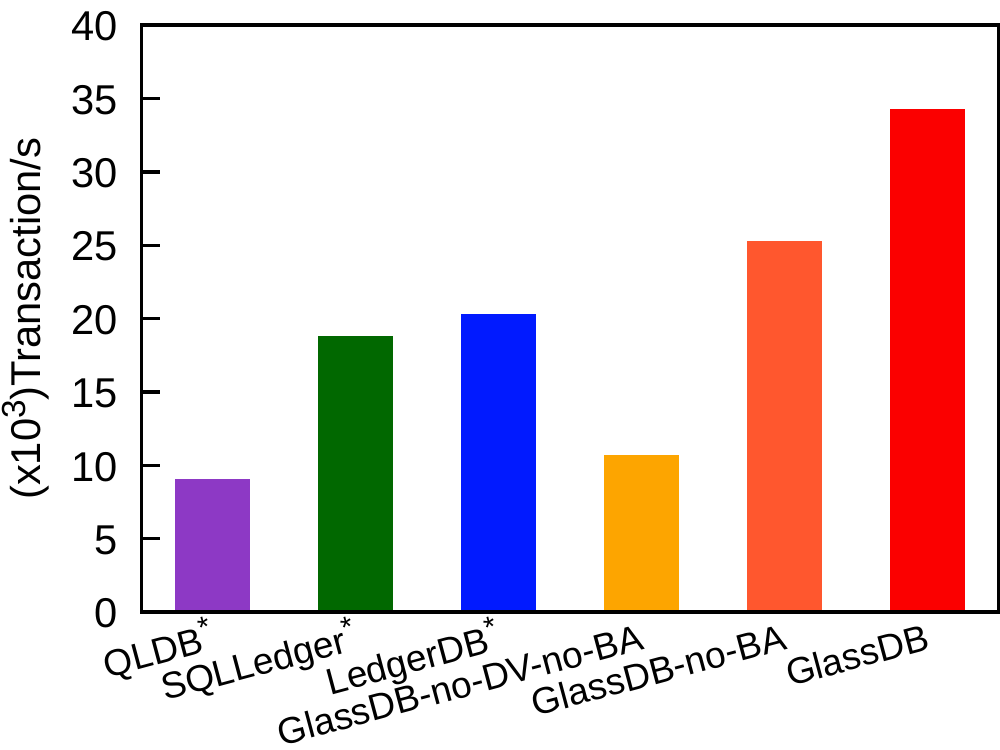}
        \caption{Impact of design choices.}
        \label{fig:exp:contribution}
    \end{minipage}
    \begin{minipage}{0.74\linewidth}
        \centering
        \subfigure[Throughput, 16 nodes]{
            \centering
            \includegraphics[height=2.9cm]{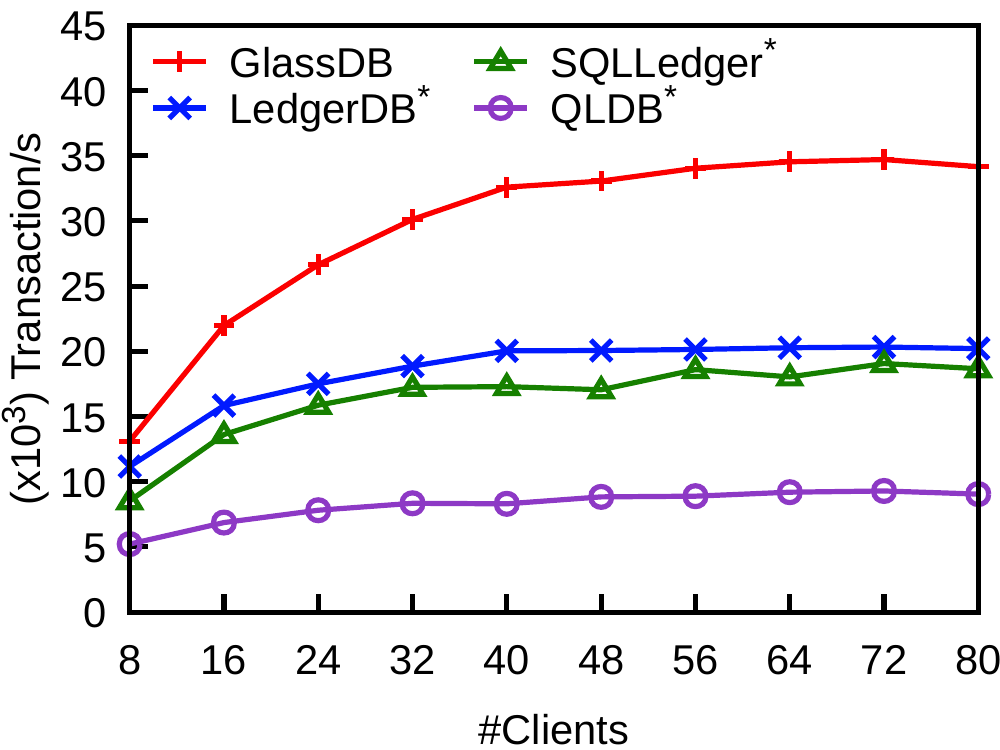}
            \label{fig:exp:ycsb_peak}
        }
        \subfigure[Throughput of varying nodes]{
            \centering
            \includegraphics[height=2.9cm]{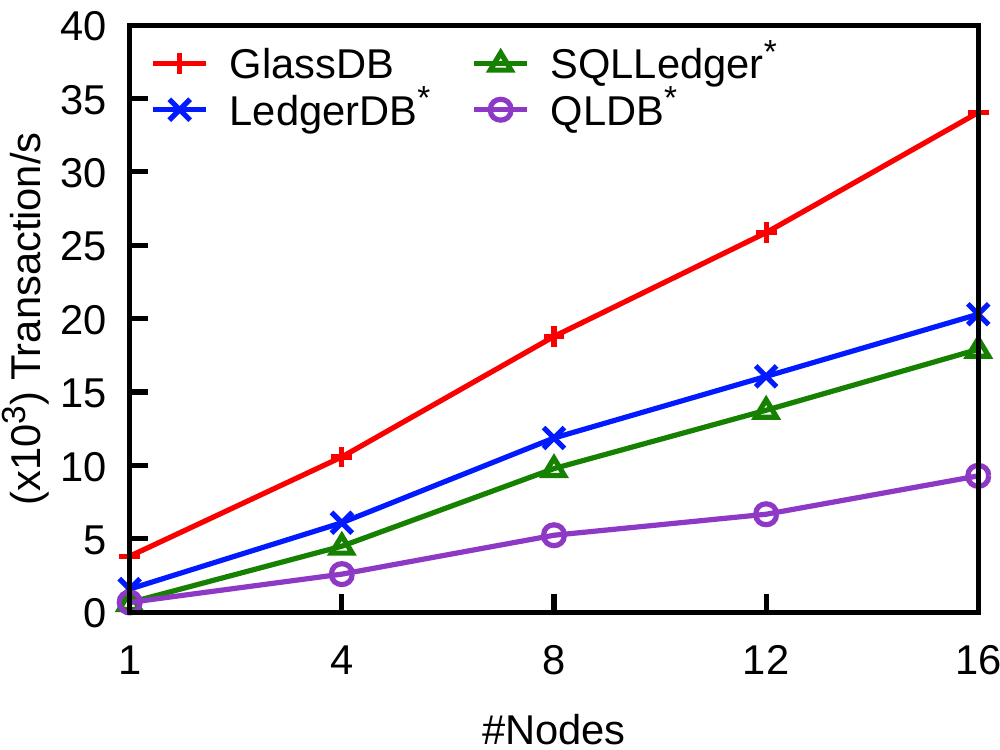}
            \label{fig:exp:ycsb_scale}
        }
        \subfigure[Read-write workload]{
            \centering
            \includegraphics[height=2.9cm]{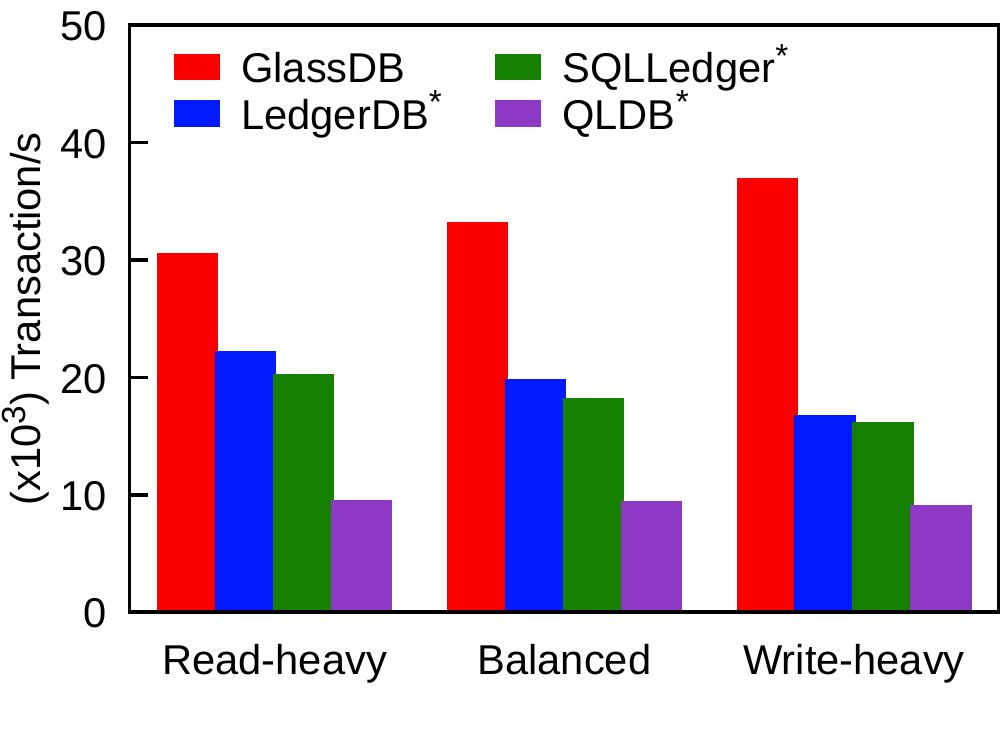}
            \label{fig:exp:ycsb_workload}
        }
        \vspace{-3mm}
        \caption{Performance for YCSB uniform workloads.}
        \label{fig:exp:ycsb}
   \end{minipage}
\vspace{-2mm}
\end{figure*}

\subsubsection{Impact of design choices}
{
To quantify the impacts of our three novel design choices, we remove these features from \systemname and compare the resulting system with the baselines.  The result is shown in Figure~\ref{fig:exp:contribution}. With only the two-level POS-tree, the system (\systemname-no-DV-no-BA) outperforms QLDB$^\ast$ by $1.2\times$. By adding deferred verification, the system (\systemname-no-BA) improves the performance by $2.4\times$, outperforming LedgerDB$^\ast$ and SQL Ledger$^\ast$, systems with deferred verification, by $1.3\times$ and $1.4\times$ respectively. By further adding batching, the throughput of the final system (\systemname) increases by another $1.3\times$.
}

\subsection{YCSB Workloads}





In the experiments, we run read-heavy, balanced, and write-heavy workloads with the number of nodes ranging from 1 to 16, the number of clients ranging from 8 to 80, and delay time ranging from 10ms to 1280ms. We first measure peak throughput by fixing the number of nodes to 16, while increasing the number of clients until the systems are saturated. Figure~\ref{fig:exp:ycsb_peak} compares the systems under the balanced uniform workload, i.e. Zipf factor is 0. \systemname outperforms
QLDB$^\ast$, LedgerDB$^\ast$, and SQL Ledger$^\ast$ by up to $3.7\times$, $1.7\times$, and $1.8\times$ respectively. \systemname, LedgerDB$^\ast$, and SQL Ledger$^\ast$ are better than QLDB$^\ast$ because they persist the authenticated data structure asynchronously, that is, they avoid updating the Merkle tree in the critical path. 
Furthermore, they all use batching that helps improve the throughput. 
\systemname's batching is more effective at reducing the overall tree heights, therefore the system is more efficient for the get-proof phase and for the verification.

\remove{Figure~\ref{fig:exp:ycsb_lat} shows the average latency at peak throughput.
\systemname has the lowest average latency, because the client gets the response as soon as the transaction is committed to the write-ahead log. The newly committed data at the server is also kept in memory to speed up queries.
The latency of LedgerDB$^\ast$ is higher because it persists the transaction blocks to  storage, and getting a key requires scanning the block.
In contrast, QLDB$^\ast$ persists the ledger synchronously, which results in the highest latency.
Next, we increase the number of server nodes from 1 to 16 and compare the peak throughput. }
Figure~\ref{fig:exp:ycsb_scale} shows that all systems scale linearly, and \systemname achieves the highest throughput. The linear scalability demonstrates that the two-phase commit's overhead is small.
\remove{To understand the improvement over the single-node setting, we record the
execution time and the number of keys processed at the server node for each transaction.
The results in Figure~\ref{fig:exp:distribution} show that the execution time
and the number of keys processed per node decrease with more nodes, which
explains the higher throughputs as the system scales.}

Figure~\ref{fig:exp:ycsb_workload} displays the throughput comparison under different workloads. 
It can be seen that \systemname consistently outperforms the baselines across all workloads. In particular, its throughput increases with a higher write ratio, because  write operations are more efficient as data is kept in memory. We note that a higher write ratio leads to more aborts and larger search space for conflicts. 
For QLDB$^\ast$, the time to update the Merkle tree is dominant, therefore the abort rate is a key factor that affects the throughput. For LedgerDB$^\ast$ and SQL Ledger$^\ast$, since the update of the Merkle tree is asynchronous, its reduction of throughput is due to higher disk I/Os and larger conflict search space. 
\remove{Figure~\ref{fig:exp:ycsb_zipf} and Figure~\ref{fig:exp:ycsb_abort} show the impact of increasing conflicts.  It can be seen that the throughputs of all systems drop as there are more aborts. Figure~\ref{fig:exp:ycsb_not_scale} illustrates how the system scales with more nodes. Unlike with uniform workloads, we observe that the throughputs scale sub-linearly, because more transactions are aborted as the number of nodes increases. }

\begin{figure*}
    \centering
    \begin{minipage}[b]{0.6\linewidth}
        \centering
        \subfigure[Throughput, 16 nodes]{
            \centering
            \includegraphics[height=3cm]{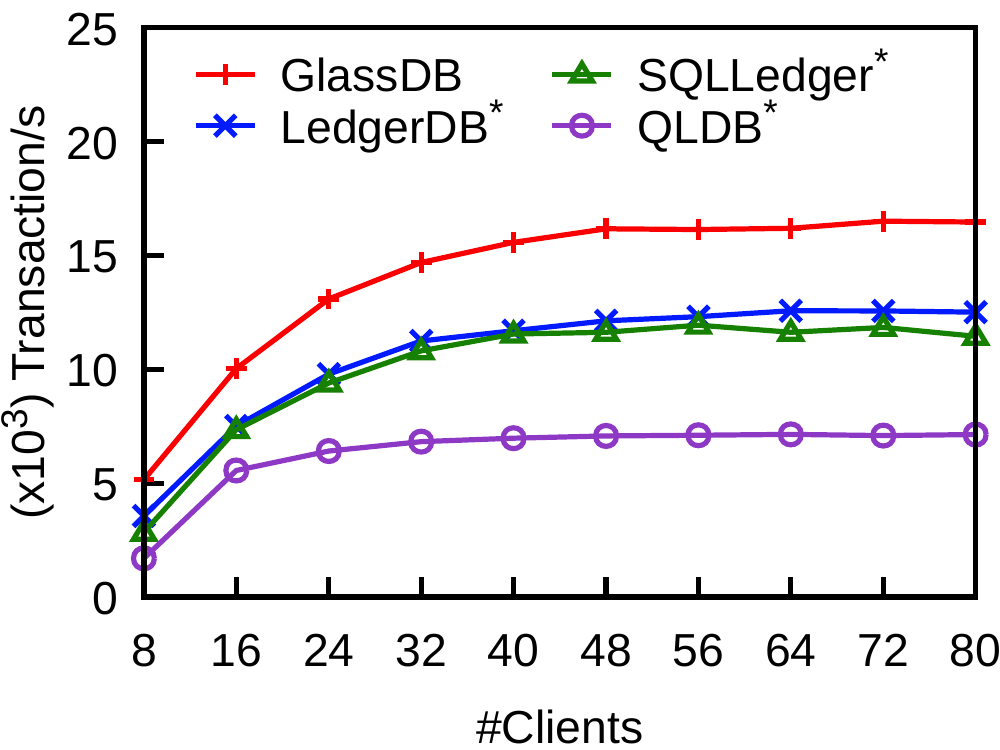}
            \label{fig:exp:tpcc_peak}
        }
        \remove{\subfigure[Throughput, varying number of nodes]{
            \centering
            \includegraphics[height=3cm]{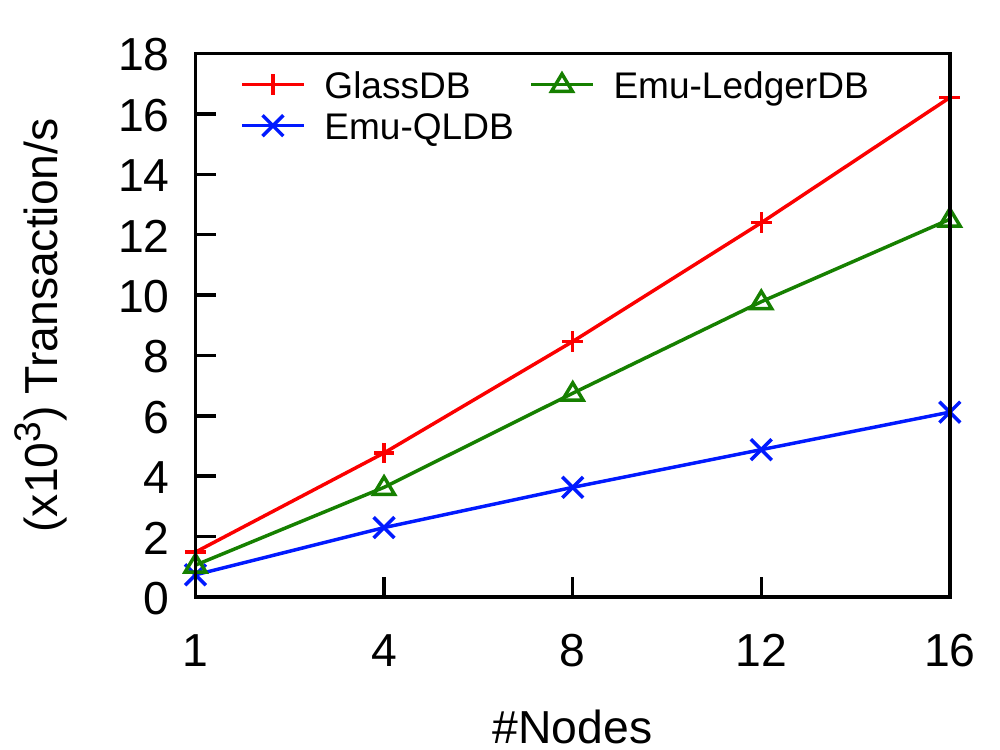}
            \label{fig:exp:tpcc_scale}
        }}
        \subfigure[Average latency breakdown]{
            \centering
            \includegraphics[height=3cm]{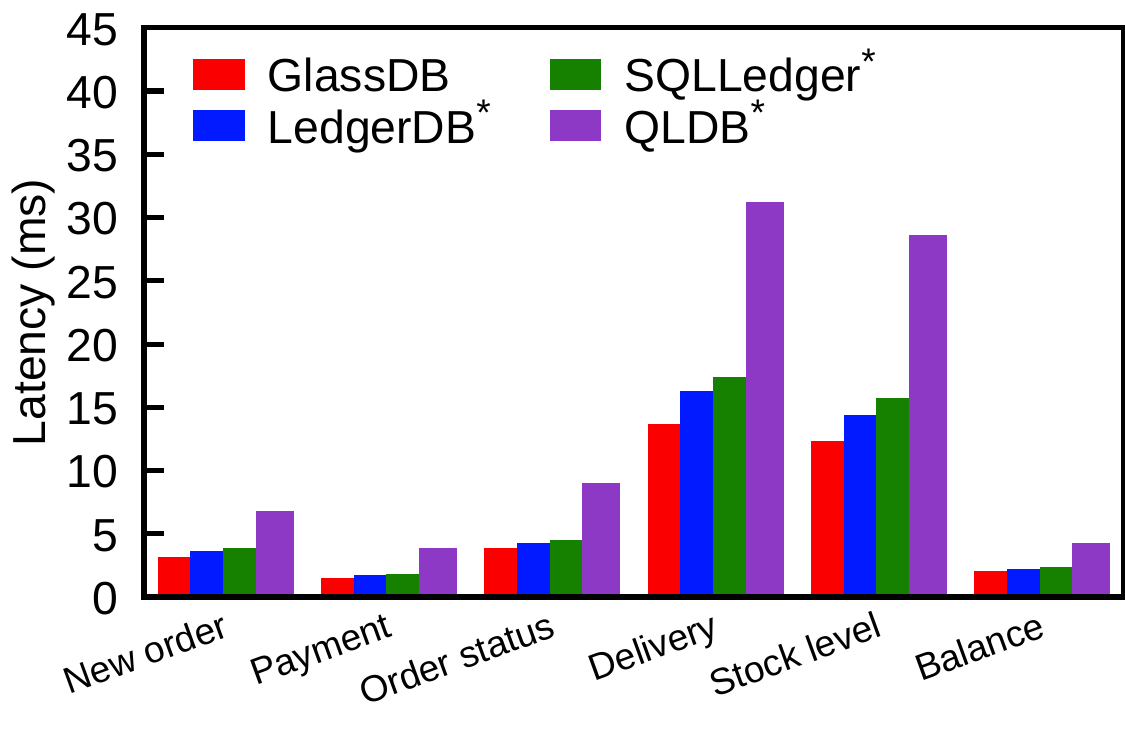}
            \label{fig:exp:tpcc_lat_op}
        }
        \caption{Performance for TPC-C workloads.}
        \label{fig:exp:dist_tpcc}
   \end{minipage}
   \begin{minipage}[b]{0.38\linewidth}
         \centering
         \includegraphics[height=3cm]{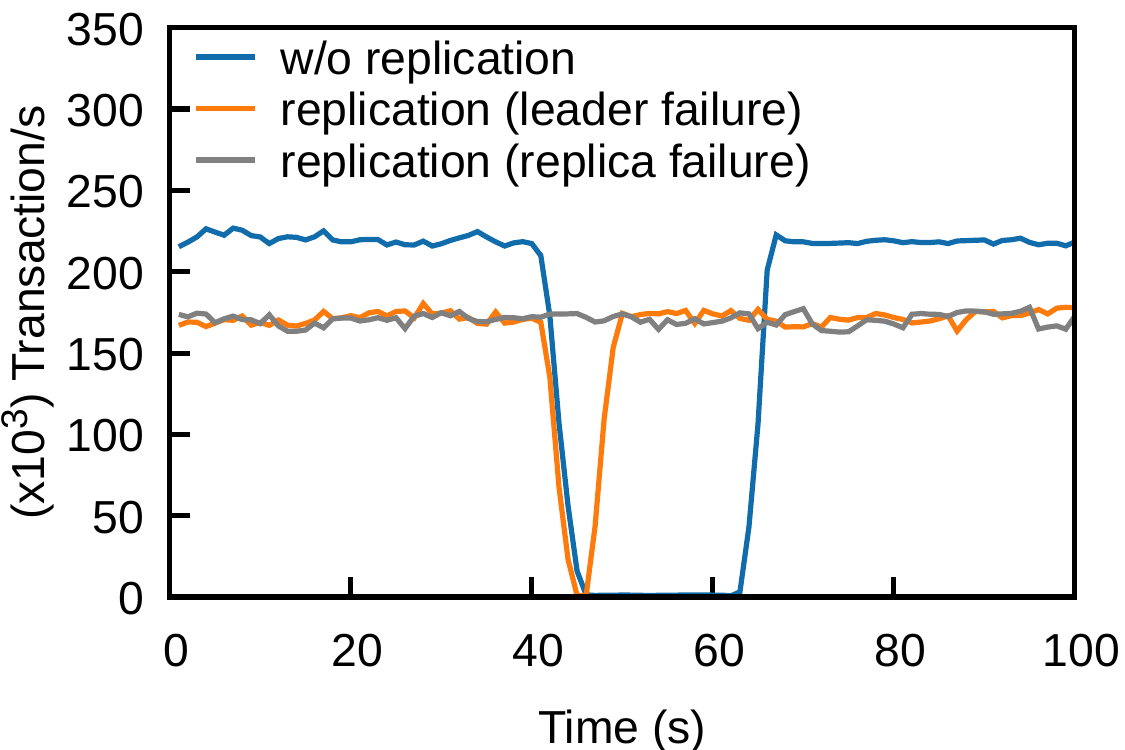}
         \vspace{6mm}
         \caption{Failure recovery.}
         \label{fig:exp:ft}
   \end{minipage}
\end{figure*}

\begin{figure*}[!t]
    \centering
    \begin{minipage}[b]{0.48\linewidth}
        \subfigure[Throughput of Workload-X]{
            \centering
            \includegraphics[height=2.9cm]{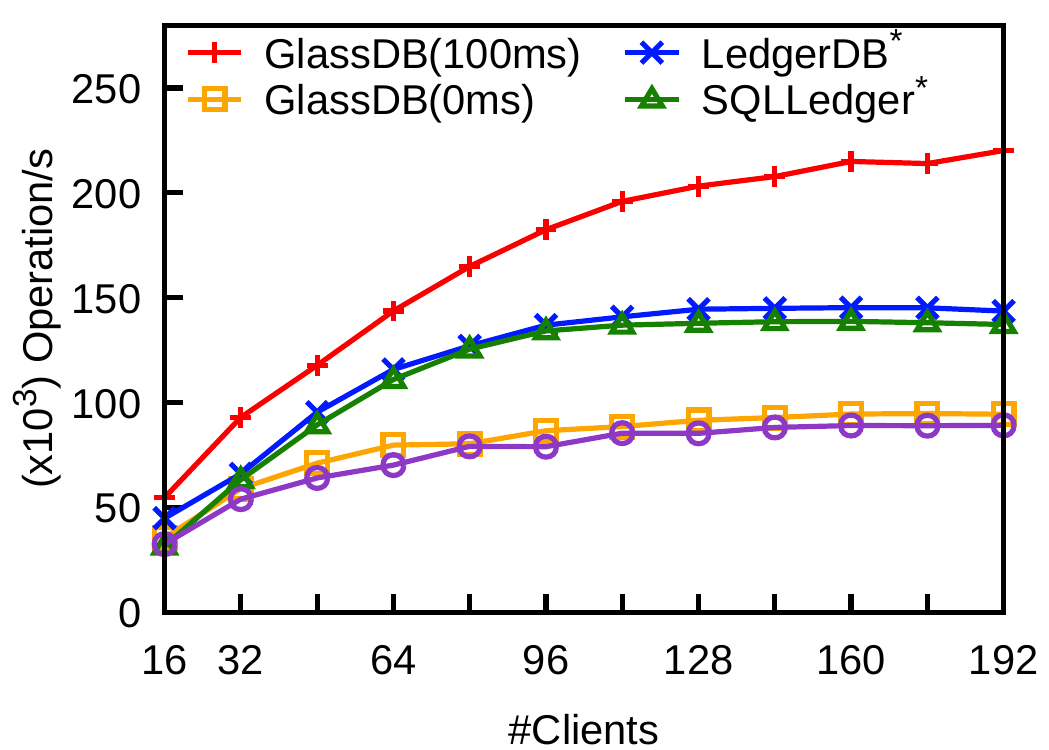}
            \label{fig:exp:dist_x}
        }
        \hspace{-3mm}
        \subfigure[Latency of Workload-X]{
            \centering
            \includegraphics[height=2.9cm]{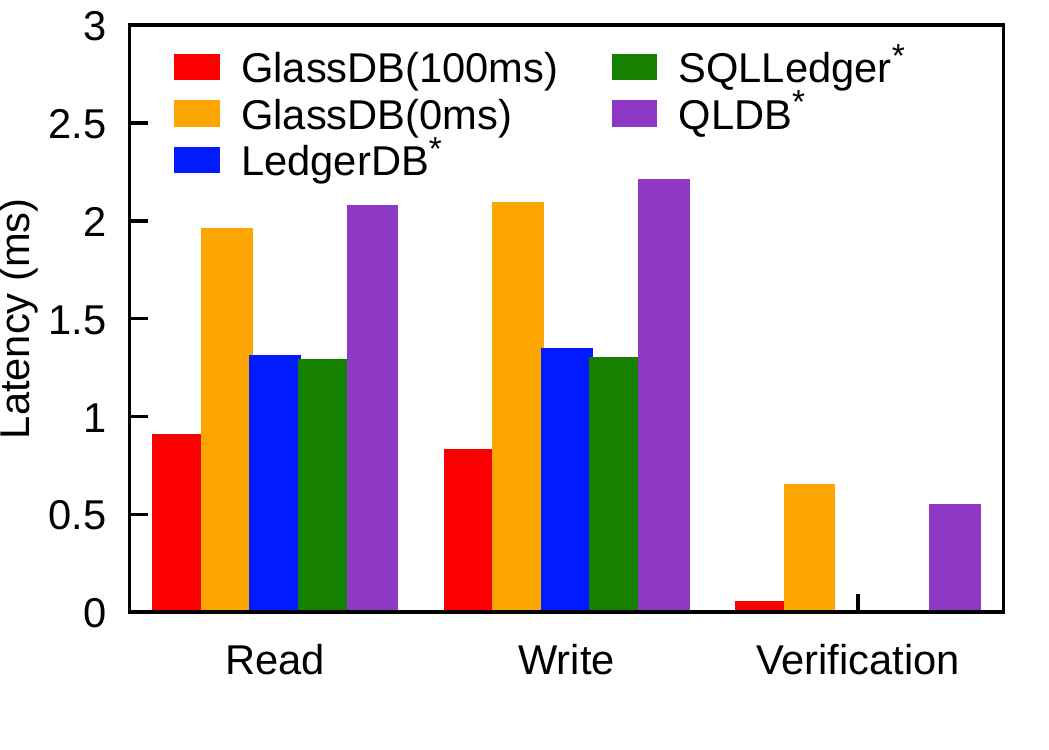}
            \label{fig:exp:dist_lat_x}
        }
        \caption{Workload-X with 16 nodes.}
    \end{minipage}
    \begin{minipage}[b]{0.24\linewidth}
        \subfigure{
            \centering
            \includegraphics[height=2.9cm]{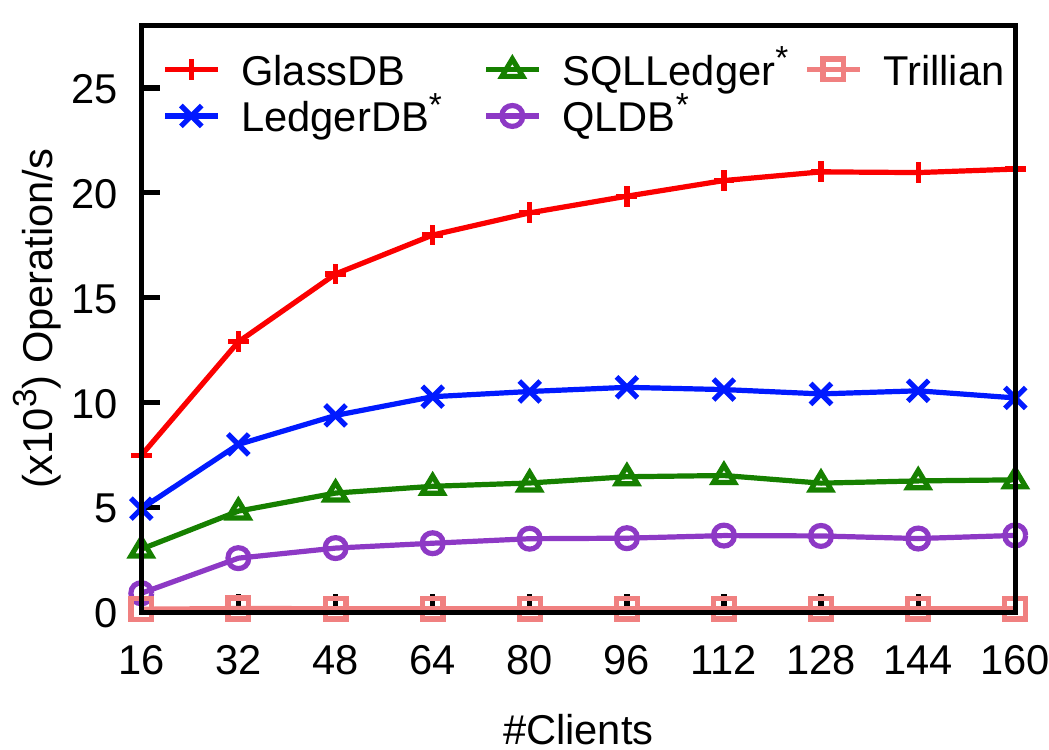}
        }
        \caption{Workload-X on a single node.}
        \label{fig:exp:singlenode}
   \end{minipage}
    \hspace{2mm}
   \begin{minipage}[b]{0.24\linewidth}
        \hspace{-3mm}
        \subfigure{
            \includegraphics[height=2.9cm]{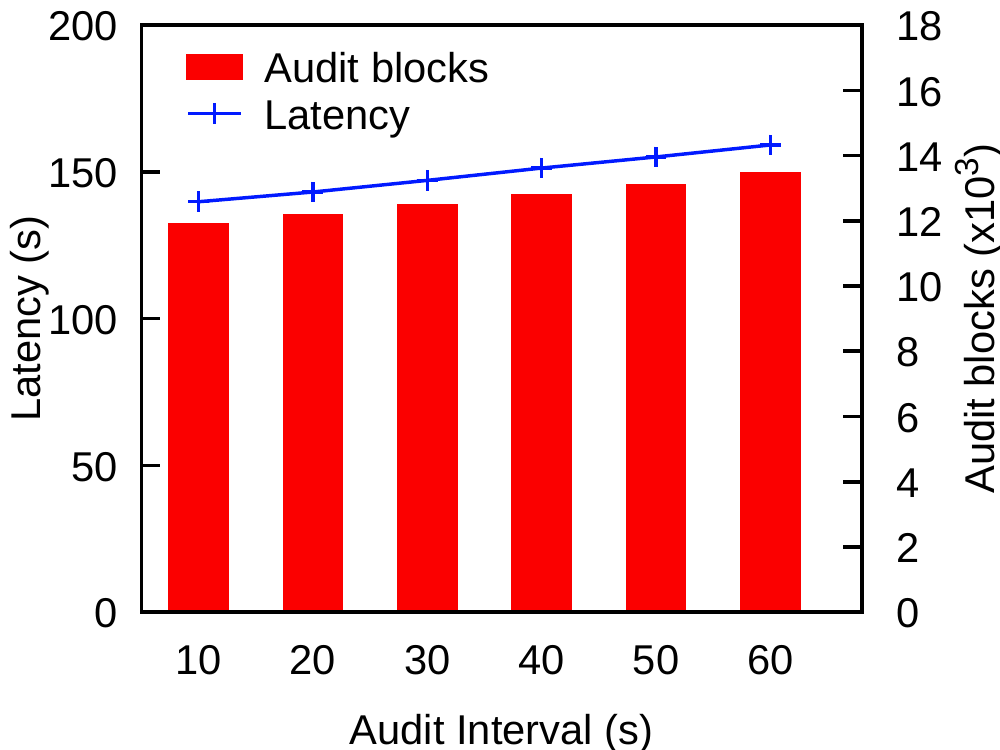}
        }
        \caption{Auditing performance.}
        \label{fig:exp:audit}
    \end{minipage}
\end{figure*}

\vspace{-2mm}
\subsection{TPC-C Workloads} 
We run a mixed workload containing all six types of TPC-C transactions mentioned in Section~\ref{subsec:tpccbench}, with new order and payment transactions accounting for 42\%, and other types accounting for 4\% each. 
We implement the tables in TPC-C on top of the key-value stores in all the systems. 
In particular, each field in a row is a data unit, and the 
key-value pair becomes <ColumnName\_PrimaryKey, FieldValue>.  We make a further optimization to
combine fields that are not frequently updated.  For example, we combine c\_first, c\_middle, and c\_last
to c\_name.

Figure \ref{fig:exp:tpcc_peak} shows the throughput with an increasing number of clients.
\systemname outperforms QLDB$^\ast$, LedgerDB$^\ast$, and SQL Ledger$^\ast$ by $2.3\times$, $1.3\times$ and $1.4\times$ respectively. 
We note that the TPC-C workload has larger and more complex transactions than YCSB workloads, therefore we observe $2.1\times$ lower throughput.
In particular, \systemname achieves peak throughput at $48$ clients, as opposed to $64$ clients for YCSB. This is because large transaction size involves more nodes and increases the overhead of coordination. 
\remove{The average latency of the systems are shown in Figure \ref{fig:exp:tpcc_lat}. The results are consistent with the throughput performance.}
Figure~\ref{fig:exp:tpcc_lat_op} shows the latency breakdown at the peak throughput for each transaction
type;
\systemname consistently has the lowest latency among all types of transactions.

\subsection{Verification Workloads}
We use workload-X as described in Section~\ref{sec:benchmark} to compare \systemname with
QLDB$^\ast$, LedgerDB$^\ast$, SQL Ledger$^\ast$, and Trillian. We omit the results for workload-Y since it displays a similar trend. We run key-value workloads, as Trillian does not support concurrent transactions.
For each verified operation, the client performs verification of the proof.



In the distributed settings with 16 nodes, 
Figure~\ref{fig:exp:dist_x} shows the throughput for workload-X with an increasing number of clients. 
\systemname achieves the highest throughput: $2.5\times$ higher than QLDB$^\ast$, $1.5\times$ higher than LedgerDB$^\ast$, and $1.6\times$ higher than SQL Ledger$^\ast$.
We evaluate the impact of deferred verification by measuring the throughput with $0$ms delay, that is, every operation is verified synchronously. Without deferred verification, the throughput is lower than LedgerDB$^\ast$ and SQL Ledger$^\ast$, and higher than QLDB$^*$. 
Figure \ref{fig:exp:dist_lat_x} shows the latency for each operation. For \systemname, LedgerDB$^\ast$, and SQL Ledger$^\ast$, which use deferred verification, we separate out the cost of verifying one key. 
\systemname outperforms the other systems in the read and write latency due to its efficient proofs (smaller proof sizes) and efficient persist phase. 
Even when combining the cost of transaction execution with that of verification, the total cost
of \systemname, LedgerDB$^\ast$, and SQL Ledger$^\ast$ are still lower than that of
QLDB$^\ast$. This is because the verification request contains multiple keys, and the
three former systems can batch multiple keys in the same proof, whereas QLDB$^\ast$ has
one proof per key.
\remove{Even with the combination of transaction execution and verification latency, the total cost
for \systemname and LedgerDB$^\ast$ are still smaller than QLDB$^\ast$, as each verification request contains 5 keys on average.}
\remove{Figure \ref{fig:exp:dist_y} and \ref{fig:exp:dist_lat_y} show similar results for workload-Y.
However, we observe verification latency for \systemname and LedgerDB$^\ast$ are higher because of larger proofs.} 

To fairly compare with Trillian, which is a single-node system that only supports key-value abstraction, we use the single-node version of \systemname, LedgerDB$^\ast$, SQL Ledger$^\ast$, and QLDB$^\ast$. The results are shown in Figure \ref{fig:exp:singlenode}. \systemname outperforms QLDB$^\ast$, LedgerDB$^\ast$, SQL Ledger$^\ast$, and Trillian by up to $5.7\times$, $2.0\times$, $3.3\times$ and two orders of magnitude, respectively. The performance gap is due to the cost of the put and get operations in Trillian being orders of magnitude more expensive. In particular, Trillian stores all data in a separate, local MySQL database instance, thus each operation incurs cross-process overheads.

Finally, we evaluate the cost of the auditing process.  We use $16$ servers with
$64$ clients, running the balanced transaction workload. After an interval, an auditor
sends $\texttt{VerifyBlock(.)}$ requests to the servers and verifies all the new
blocks created during the interval. Figure~\ref{fig:exp:audit} shows the
auditing costs with varying intervals from 10s to 60s. Both the latency for
verifying the new blocks, and the number of the new blocks grow almost linearly
with the audit interval. This is because more blocks are created during a
longer interval, and it takes a roughly constant time to verify
each block. We remark that the auditing process is expensive, especially when
the rate of block creation is high. However, it can be done off the critical path,
and is amenable to distributed processing.

\subsection{Failure Recovery}

In this section, we test the impact of crash failures and the recovery process of \systemname. We compare the two implementations, i.e., the failure recovery with respect to 2PC and fault tolerance with replication, as described in Section \ref{subsec:ft}. We set the replica group size to be 3 to tolerate one node failure for the replication setting. The experiment is conducted with 16 nodes and 160 clients. After the performance of the system becomes stable, we let the system continues running for 40 seconds. Next, we kill one node and reboot it after 20 seconds. After that, we let the system run for another 40 seconds. We take measurements of the throughput for every second. The results are shown in Figure ~\ref{fig:exp:ft}. In the normal scenario where no crash failure occurs, replication contributes to around 22\% overhead. When one node crashes, \systemname without replication has to abort all transactions accessing keys in the partition hosted by the failed node, therefore, has low throughput until the crashed node is brought back at 60 seconds.
For \systemname with replication, in the case of a leader failure, the system encounters a temporary low throughput because of leader election, syncing of the states, and transaction aborts due to timeouts. It takes around 7 seconds to recover to peak throughput and continues to work as normal. In the case of a replica failure, the system will continue to process the transactions at the peak throughput.

\section{Related Work}
\label{sec:related}

\noindent
{\bf Verifiable OLAP databases.} 
Zhang et al.~\cite{ZhangGKPP17} propose interactive protocols for verifiable SQL queries. However, their techniques rely on expensive cryptographic primitives
Systems that use trusted hardware include EnclaveDB~\cite{enclavedb}, Opaque~\cite{opaque}, and
ObliDB~\cite{oblidb}, and they support full-fledged SQL query execution inside trusted enclaves.
VeritasDB~\cite{veritasdb18} and Concerto~\cite{concerto17} leverage trusted hardware to ensure the integrity of key-value operations.
VeriDB~\cite{veridb} extends Concerto to supports general SQL queries.
All of these systems make a strong security assumption on the availability and security of the trusted hardware.


\noindent
{\bf Authenticated data structure.} Li et al.~\cite{ads} propose multiple 
index structures based on Merkle tree and B$^+$-tree.
IntegriDB~\cite{integridb} proposes efficient authenticated
data structures that support a wide range of queries such as join and aggregates. We note that these data
structures do not guarantee the integrity of the data history.


\noindent
{\bf Blockchain databases.} 
Veritas~\cite{veritas} proposes a verifiable table abstraction, by storing transaction logs on a blockchain.
vChain~\cite{vchain19} and FalconDB~\cite{falcondb} combine authenticated data structures
with blockchain, by storing digests of the authenticated index structures in the blockchain. 
LineageChain~\cite{lineagechain} enables efficient access of data provenance information for Hyperledger.
The main disadvantage of blockchain-based systems is that they have poor performance. 
The main disadvantage of blockchain-based systems is that they have poor performance.

\section{Conclusions}
\label{sec:conclusion}
In this paper, we described the design space of verifiable ledger databases. We designed and implemented
\systemname that addresses the limitations of existing systems. \systemname supports transactions, has  
efficient proofs, and high performance. We evaluated our system against three baselines, using new
benchmarks supporting verification workloads.
The results show \systemname significantly outperforms the baselines.


\bibliographystyle{ACM-Reference-Format}
\balance
\bibliography{main-bibliography}

\end{document}